\documentclass[preprint2,useAMS,usenatbib]{aastex}
\usepackage{graphicx,amsmath}
\pdfoutput=1

\usepackage{capt-of}

\newcommand{\sdssnearby}{SDSS$_{\text{nearby}}$}
\newcommand{\hi}{\ion{H}{1}} 
\shortauthors{Moorman et al.}
\shorttitle{Void Galaxy LFs}

\begin{document}

\title{The Optical Luminosity Function of Void Galaxies in the SDSS and ALFALFA Surveys} 
\author{Crystal M. Moorman, Michael S. Vogeley}
\affil{Department of Physics, Drexel University, 3141 Chestnut Street, Philadelphia, PA 19104}
\email{crystal.m.moorman@drexel.edu}
\author{Fiona Hoyle}
\affil{Pontifica Universidad Catolica de Ecuador, 12 de Octubre 1076 y Roca, Quito, Ecuador}
\author{Danny C. Pan}
\affil{Shanghai Astronomical Observatory, Shanghai, China, 200030}
\and
\author{Martha P. Haynes, Riccardo Giovanelli}
\affil{Center for Radiophysics and Space Research, Space Sciences Building, Cornell University Ithaca, NY 14853}

\begin{abstract}
We measure the r-band galaxy luminosity function (LF) across environments over the 
redshift range $0<z<0.107$ using the SDSS. We divide our sample into galaxies residing 
in large scale voids (void galaxies) and those residing in denser regions (wall galaxies). 
The best fitting Schechter parameters for void galaxies are: 
$\log\Phi^*= -3.40 \pm 0.03$ $\log($Mpc$^{-3})$, 
$M^*= -19.88\pm 0.05$, and $\alpha= -1.20\pm 0.02$. For wall galaxies, 
the best fitting parameters are: $\log\Phi^*=-2.86 \pm 0.02$ $\log($Mpc$^{-3})$, 
$M^*= -20.80\pm 0.03$, and $\alpha= -1.16\pm 0.01$.
We find a shift in the characteristic magnitude, $M^*$, towards fainter magnitudes for void galaxies 
and find no significant difference between the faint-end slopes of the void and wall galaxy LFs. 
We investigate how low surface brightness selections effects can affect the galaxy LF. 
To attempt to examine a sample of galaxies that is relatively free of surface brightness selection effects, 
we compute the optical galaxy LF of galaxies detected by the blind \hi\ survey, ALFALFA. 
We find that the global LF of the ALFALFA sample is not well fit by a Schechter function, 
because of the presence of a wide dip in the LF around $M_r=-18$ and an upturn at fainter magnitudes ($\alpha\sim-1.47$).   
We compare the \hi\ selected r-band LF to various LFs of optically selected populations to determine where the \hi\ 
selected optical LF obtains its shape. We find that sample selection plays a large role in determining the shape of the LF.  
\end{abstract}


\section{INTRODUCTION}  
A critical measurement of the distribution of galaxies, 
that may be compared to galaxy formation and evolution models, is the 
galaxy luminosity function (LF): the number of galaxies per volume per luminosity. 
The shape of a LF is typically fit by a \cite{Schechter1976} function described as 
a power law with faint-end slope, $\alpha$, 
and an exponential drop-off at the characteristic magnitude, $M^*$. 
With the advent of deep, wide optical surveys, we now have sufficiently large samples of 
galaxies that allow us to study the LF of complete samples of galaxies across environments, 
colors, and morphological types.  

Deep redshift surveys have allowed measurements of the evolution of the LF with cosmic time 
\citep{Ramos2011,Loveday2012,McNaught2014,Martinet2014redshift}. 
Measuring how the LF changes with galaxy morphology and color can support or reject  
theories on galaxy formation and evolution. For instance, the LF results of \cite{Driver2007} 
suggest that galaxies begin their lives with a pseudo-bulge-like structure and evolve into 
disc galaxies before becoming classic elliptical galaxies. 
\cite{Madgwick2002},\cite{Driver2007},\cite{Yang2009}, and \cite{Tempel2011} find that the galaxy LF 
varies with both color and morphological type. Early-type galaxies typically have larger 
characteristic magnitudes and flatter faint-end slopes than late-type galaxies. Similarly, red galaxies tend to 
have flatter faint-end slopes and slightly larger characteristic magnitudes than blue galaxies. 
They also find that the bright end of the LF is characteristically determined by red and elliptical galaxies, 
while the faint end is dominated by blue and spiral galaxies. 

Determining how LFs split based on color and type vary with environment gives us insight as to how 
large-scale structure affects the evolution of galaxies. 
For example, \cite{Tempel2011} find that the shape of the spiral galaxy LF is independent of environment. 
This implies that spiral galaxies in voids evolve no differently than spiral galaxies in denser regions. 
They also find that the faint-end of the elliptical galaxy LF steepens with increasing density. 
This effect is likely due to the presence of satellite galaxies that formed in the early Universe 
and have since been tidally stripped quenching star formation at later times. 
However, these authors only probe the environmental effects on the LF of galaxies down to $M_r=-17$. 

Previous measurements of the galaxy LF dependence on large-scale structure range from 
void regions \citep{GroginGeller99,Hoyle2005,Tempel2011}, 
to groups \citep{Eke2004,Yang2009}, 
to clusters \citep{DePropris2003,Durret2011IAU,McNaught2014,Martinet2014cluster}. 
The environmental effects on the characteristic magnitude remain consistent 
across the literature: $M^*$ becomes fainter with decreasing large-scale density. 
The environmental dependence of $\alpha$ varies in the literature depending on the survey, 
redshift limits, magnitude limits, and methods used. 

Previous estimates of the best fit Schechter function faint-end slopes of the galaxy LF in voids 
are not well constrained. Estimates range from 
$\alpha=-1.18\pm0.13$ for a small sample of galaxies brighter than $M_r=-14.5$ 
\citep{Hoyle2005} to 
$\alpha=-0.98\pm0.02$ for galaxies brighter than $M_r=-17$ 
\citep{Tempel2011} to 
$\alpha=-1.36\pm0.05$ for galaxies brighter than $M_r=-17$ 
\citep{McNaught2014}. 
In the dwarf ($M_r>-17$) regime, \cite{Hoyle2005} find that the faint-end slope remains the same between 
void galaxies and those in denser regions.
The best fit LF faint-end slopes of galaxy groups vary from 
$\alpha\sim-1.0$ to $-1.2$ where the faint-end slope steepens 
with increasing group mass \citep{Eke2004,Tully2005,Yang2009}. 
As for dwarf galaxies in groups, \cite{Tully2005} finds that more dynamically evolved regions 
have steeper faint-end slopes. 
The faint-end slopes of composite LFs for clusters yield slopes ranging from $\alpha\sim-1.25$ to $-1.4$ 
\citep{Valotto1997, DePropris2003, Durret2011IAU, McNaught2014, Martinet2014cluster}. 
Additionally, \cite{Tempel2011} find that elliptical galaxies dominate the LF in high-density regions. 
This corroborates the luminosity-morphology-density relation of \cite{Park2007} 
and the work of \cite{Hoyle2012} who find that galaxies in 
the most underdense voids are primarily bluer, late-type galaxies. 
The overall trend for nearby groups and clusters seems to be that the faint end slope steepens with increasing density. 
Where the void galaxy LF faint-end slope fits in to this trend is currently uncertain. 
To date, studies using large ($>10^4$) sample sizes have not probed the void LF down to the dwarf regime ($M_r>-17$). 

In this work, we will investigate the optical LF of void galaxies down to $M_r=-13$. 
The dwarf faintest end of the LF that we probe could be plagued by low-surface brightness selection effects.  
\cite{Blanton2005ELL} estimate the LF of extremely low luminosity galaxies, 
but do not explore the environmental effects. These authors also examine the effects of 
low-surface brightness selection in the extremely low luminosity sample and find that 
accounting for low surface brightness selection effects steepens the global LF faint-end slope 
from $\alpha\sim-1.3$ to $\alpha\sim-1.5$. 

To avoid the optical selection bias against low surface brightness galaxies, 
we would like to determine the effects on the optical LF of a sample of \hi\ selected galaxies. 
\cite{Zwaan2001} estimate the LF using an \hi\ selected sample 
from the Arecibo \hi\ Strip Survey (AHISS), 
an approach that is free from optical selection biases. These authors find the 
LF of AHISS detections are in good agreement with those of late-type galaxies. 
However, \cite{Zwaan2001} have a sample size of only 60 \hi\ detections. 
For the first time, we have a statistically significant sample of matched optical and \hi\ data 
with which we can estimate the optical galaxy LF. 

Blind \hi\ surveys typically detect bluer galaxies 
than optically selected surveys with large values of 
the inverse concentration index which is characteristic of late-type galaxies.
Given the known similarities between the void galaxies \citep{Hoyle2012,CrotonFarrar2008} and 
\hi\ selected galaxies \citep{Toribio2009,Huang2012} 
(e.g. blue, late-type, high-specific star formation rates), 
we expect to find a larger fraction of void galaxies in our \hi\ selected sample \citep{Moorman2014} and 
less environmental dependence on the optical LF of \hi\ selected samples 
than found for optically selected samples. 
Namely, a smaller shift in characteristic magnitudes and similar faint-end slope between void and wall LFs. 
Thus, we will also test for environmental dependencies on the optical LF of \hi\ selected galaxies.

$\Lambda$CDM simulations predict a steep ($\alpha\sim-1.8$) power law slope at the faint end 
of the dark matter halo mass function \citep{MathisWhite2002,Klypin2011}. 
Halo occupation distribution models provide a statistical description of 
the number of luminous galaxies that occupy a halo of given mass: 
massive halos host a ``central'' galaxies as well as less luminous ``satellites,'' 
while very low mass halos may host, at most, one central galaxy.  
$\Lambda$CDM simulations predict that the characteristic mass of the dark matter halo mass function 
shifts toward lower mass in low density regions \citep{Goldberg2005}. 
Together, these results imply that faint galaxies in voids are likely to be``central'' galaxies in their own halos, 
while faint galaxies in denser regions are likely to be satellites of more luminous galaxies. 
Additionally, the hydrodynamic simulations used in \cite{Sawala2014} which were tailored to match the conditions of the 
Local Group imply that luminous, low-mass haloes near ``central'' haloes were probably once larger haloes that 
formed stars in the early Universe and have since been tidally stripped by neighboring haloes. 
These results imply a relatively steep faint-end slope in dense regions. 
Other simulations predict varying environmental effects on the faint-end slope of the galaxy LF. 
For instance, \cite{MathisWhite2002} predict a steeper faint-end slope in low-density regions, 
while \cite{Cui2011} predict that the faint-end slope in low-density regions is 
similar to that of high-density regions. 
Therefore, accurately measuring the shape of the faint-end of the LF provides strong 
constraints for formation models of dwarf galaxies. 

The observed galaxy LF slope is significantly shallower ($\alpha\sim-1.3$) 
than the predicted low-mass halo slope. 
For theory to match observations, the incorporation of feedback and photoionization effects 
is required in simulations to suppress star formation in galaxies at both the faint and bright ends. 
Current simulations (e.g. the Millennium Simulation \cite{Springel2005MS}) have 
incorporated star formation quenching effects such that the 
outcome of the predicted LF matches current measurements of the observed LF. 
However, if the current measurements are underestimating the ``true'' faint-end slope of the galaxy LF 
as predicted by \cite{Blanton2005ELL}, then the simulations may need to scale back 
the effects of feedback and photoionization to account for the presence of LSB galaxies. 

In this paper, we present the environmental effects on the LF of optically selected 
galaxies from the SDSS DR7 and \hi\ selected galaxies from the 
ALFALFA Survey. 
In Section \ref{sec:Data} we briefly discuss the data, void identification method, and 
compare the properties of \hi\ and optically selected samples. 
In Section \ref{sec:LF_meth} we discuss the methods used in this work. 
We present the optical LF of void and wall galaxies from 
SDSS DR7 galaxies as well as the void and wall LFs of 
\hi\ detections from the ALFALFA Survey 
in Section \ref{sec:results}. 
Here, we also discuss any differences between the optical LFs 
of \hi\ and optically selected galaxies. We discuss the conclusions 
of our work in Section \ref{sec:LF_conc}.
Throughout this work, we assume $\Omega_m=0.26$ 
and $\Omega_{\Lambda}=0.74$ when calculating comoving coordinates and absolute magnitudes.

\section{DATA} 
\label{sec:Data}
\subsection{SDSS DR7}
\label{subsec:SDSS_data}
The Sloan Digital Sky Survey (SDSS) 
\citep{Fukugita1996, Gunn1998, Lupton2001,Strauss2002,Blanton2003SpectraSDSS} 
is a wide-field multi-filter imaging 
and spectroscopic survey covering a quarter of the
northern Galactic Hemisphere in the five band SDSS system-$u,g,r,i$, 
and $z$. 
Once each image is classified, follow up spectroscopy is done on 
galaxies with r band magnitude $r < 17.77$.
Spectra obtained through the SDSS are taken 
using two double fiber-fed spectrographs and fiber plug plates covering a portion 
of the sky $1.49^{\circ}$ in radius with minimum fiber separation of 55 arcseconds. 

For the optical data in this work, we utilize the  
Korea Institute for Advanced Study Value-Added Galaxy
Catalog (KIAS-VAGC) of \cite{Choi2010}. 
The KIAS-VAGC catalog is based on the SDSS Data Release 7 
(DR7) \citep{Abazajian2009} spectroscopic targets in the main galaxy sample 
and contains 707,817 galaxies.

\subsection{ALFALFA}
\label{subsec:ALF_data}
The Arecibo Legacy Fast ALFA (ALFALFA) Survey \citep{ALFALFAII,ALFALFAI} is a large-area, blind 
extragalactic \hi\ survey with sensitivity limits allowing for the detection of galaxies
with \hi\ masses down to $M_{HI}=10^8 M_{\odot}$ out to 40 Mpc. 
The most recent release of the ALFALFA Survey \citep[$\alpha.40$;][]{Haynes2011}, covers $\sim$2800 deg$^2$ 
across two regions in the northern Galactic Hemisphere, called the Spring Sky, 
($07^h30^m<$R.A.$<16^h30^m$, $04^{\circ}<$ Dec
$<16^{\circ}$ and  $24^{\circ}<$ Dec $<28^{\circ}$), and two in the 
southern Galactic Hemisphere, called the Fall Sky, 
($22^h<$R.A.$<03^h$, $14^{\circ}<$ Dec
$<16^{\circ}$ and  $24^{\circ}<$ Dec $<32^{\circ}$). 
The \hi\ detections in this catalog are categorized as either Code 1, 2, or 9.
Code 1 objects are reliable detections with high S/N ($>6.5$);
Code 2 objects have S/N$<6.5$ and coincide with optical counterparts with known 
redshift similar to the \hi\ detected redshift; and Code 9 objects are 
high velocity clouds. 

To obtain optical properties of the \hi\ sources, we use the 
cross-reference catalog of 12,468 ALFALFA \hi\ sources with the most probable 
optical counterpart from the SDSS DR7 supplied by \cite{Haynes2011}.  
Because we are interested in each galaxy's environment, we must limit our sample to objects found in the 
region accessible to the DR7 void catalog of \cite{Pan2012}, the NGC region. 
We limit our sample to Code 1 detections within $z\le0.05$, 
due to radio frequency interference beyond this redshift range. 

For our detections, we query the SDSS DR7 database to obtain 
all information needed for the analysis in this work, such as color, 
inverse concentration index, absolute magnitude, and surface brightness. 
Because we compare optically selected galaxy LFs to \hi\ selected galaxy LFs, 
we must remain consistent in how we determine absolute magnitudes, $M_r$. 
Therefore, we K-correct the magnitudes and 
band-shift each \hi\ source's $M_r$ to $z=0.1$ using 
K-correct Version 4.4 \citep{Blanton_kcorrect} as done in the KIAS-VAGC. 
Because not all ALFALFA detections have optical spectroscopy, we adopt the 21 cm redshift for 
determining each galaxy's comoving distance and $r$-band absolute magnitude. 
Approximately 200 of $\sim$7000 of the galaxies in the ALFALFA sample 
are matched to SDSS galaxies for which SDSS spectra were not taken. 
The differences between optical and \hi\ redshifts are typically less than 50 kpc s$^{-1}$. 
Only a handful of galaxies have HI to optical redshift differences in the range 50 kpc s$^{-1}$--200kpc s$^{-1}$. 
These differences are consistent with 
the velocity widths of $M^*$ galaxies; 
thus the effect on the faint-end slope of the LF will be negligible.

\subsection{Creating the Void Samples}
We classify all of our galaxies as void or wall detections 
by comparing the comoving coordinates of each to 
the void catalog of \cite{Pan2012}. This void catalog uses the galaxy-based 
void finding algorithm of \cite{HoyleVogeley2002}, VoidFinder (also, see \cite{ElAdPiran1997}). 
We briefly discuss the algorithm here: 
In a map of SDSS DR7 galaxies with $M_r<-20.1$ in the Northern Galactic Hemisphere, 
we calculate each galaxy's third nearest neighbor. If the third nearest neighbor is 
at least 7h$^{-1}$Mpc away, the galaxy is removed from the map and is considered a ``potential void galaxy.'' 
Once all ``potential void galaxies'' have been removed, VoidFinder grows spheres in the empty regions of the map.  
If a sphere has a minimum radius of 10h$^{-1}$Mpc and lies completely within the survey mask, 
it is considered a true void. We then replace the ``potential void galaxies'' back into the map. 
if these ``potential void galaxies'' lie within one of the true void spheres we classify it as a ``void galaxy''; 
otherwise, we classify it as a ``wall galaxy.'' 
We compare the locations of all of our galaxies against the locations of the void spheres. 
If a galaxy lies within a large-scale void in this catalog, we classify the galaxy as a ``void galaxy''; 
otherwise, we classify the galaxy as a ``wall galaxy.'' 
The void catalog contains 1054 voids, the largest of which are $\sim$30h$^{-1}$Mpc in effective radius. 
The median effective radius of all voids within the catalog is $\sim$17h$^{-1}$Mpc. 
See \cite{Pan2012} for further details regarding the statistical properties of the voids in SDSS DR7. 
Because we require each sphere to lie completely within the mask, we risk misclassifying 
true void galaxies along the edges as wall galaxies. 
From the optically selected sample, we identify $75,063$ ($21\%$) void galaxies 
and $274,436$ ($77\%$) wall galaxies within $z<0.107$.
The remaining galaxies (2\%) lie along the survey edges and are excluded due to the 
potential misclassification issues.

Similarly, for the \hi\ selected sample, we positionally compare each \hi\ detection's 
comoving coordinates to the void catalog. We identify 
$2,611$ ($35\%$) void detections, $4,566$ ($60\%$) wall detections, 
and 390 ($5\%$) detections located near the survey edges.

For the sake of comparing the effects of the selection biases 
between the SDSS and ALFALFA samples, we make an optically selected subsample 
which we call \sdssnearby.
For \sdssnearby, we limit ourselves to galaxies within $z<0.05$.   
We further limit the sample to galaxies 
within $07^h30^m<$R.A.$<16^h30^m$, $04^{\circ}<$ Dec $<16^{\circ}$ 
and  $24^{\circ}<$ Dec $<28^{\circ}$ to ensure we only 
use galaxies within the same volume as our ALFALFA sample. 
We compare the magnitude-limited $(r<17.6)$ \sdssnearby\ galaxy locations 
to the void catalog of \cite{Pan2012}. We identify 7,058 (25\%) void galaxies, 
20,148 (73\%) wall galaxies, and 537 (2\%) galaxies living near the survey edges. 
The reader should keep in mind that a ``void'' sample from an \hi\ survey and 
a ``void'' sample obtained from an optical survey contain fundamentally different galaxies. 

\subsection{Comparing The \hi\ and Optically Selected Samples}
\label{subsec:compare_hi_opt_selected_gals}
\begin{figure}[h]
  \begin{center}
    \includegraphics[scale=0.45,trim = 0.75cm 0.25cm 1.5cm 1.25cm, clip=True]{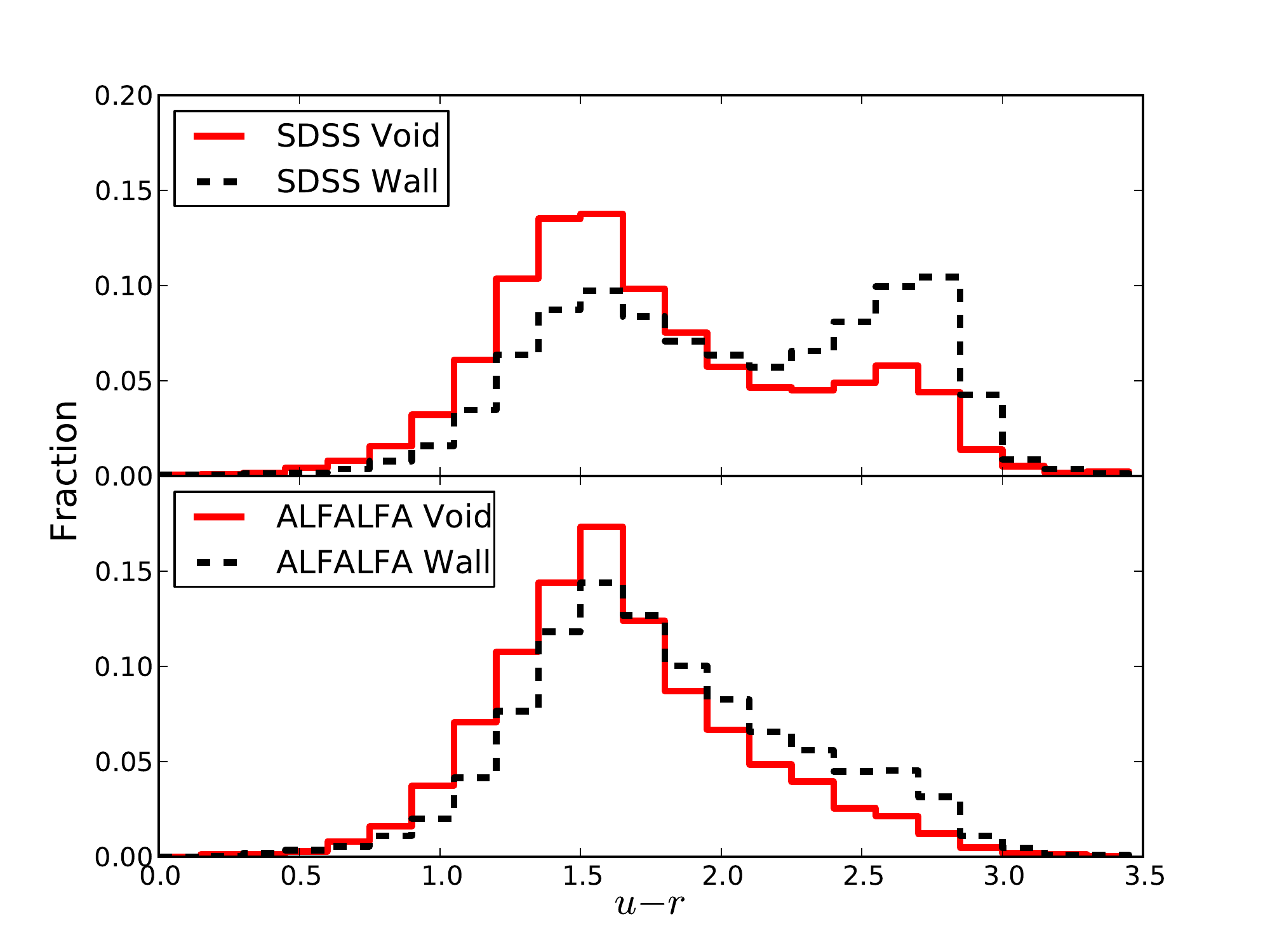}
  \end{center}
  \caption{Color distribution of \sdssnearby\ (upper) and ALFALFA (lower) 
  void (red) and wall (black) galaxies. 
  The \sdssnearby\ sample contains 7,058 void galaxies 
  and 20,148 wall galaxies. The ALFALFA sample contains 2,611 void detections and 4,566 wall detections. 
  Both SDSS distributions appear bi-modal, 
  with a less prevalent red sequence peak in the void distribution 
  as evident in the results of Hoyle et al. (2012). 
  The ALFALFA color distributions do not appear to be as 
  bimodal as the SDSS distributions do, because 
  \hi\ selection is biased against luminous, red galaxies. 
  We see a hint of the red sequence in the \hi\ selected wall distribution, 
  but the void distribution appears to follow a skewed unimodal distribution originating 
  from the blue cloud. For both surveys, void galaxies are generally bluer than wall galaxies.
    }
  \label{fig:frac_color}
\end{figure}
\begin{figure}[h]
  \begin{center}
    \includegraphics[scale=0.45,trim = 0.75cm 0.25cm 1.5cm 1.25cm, clip=True]{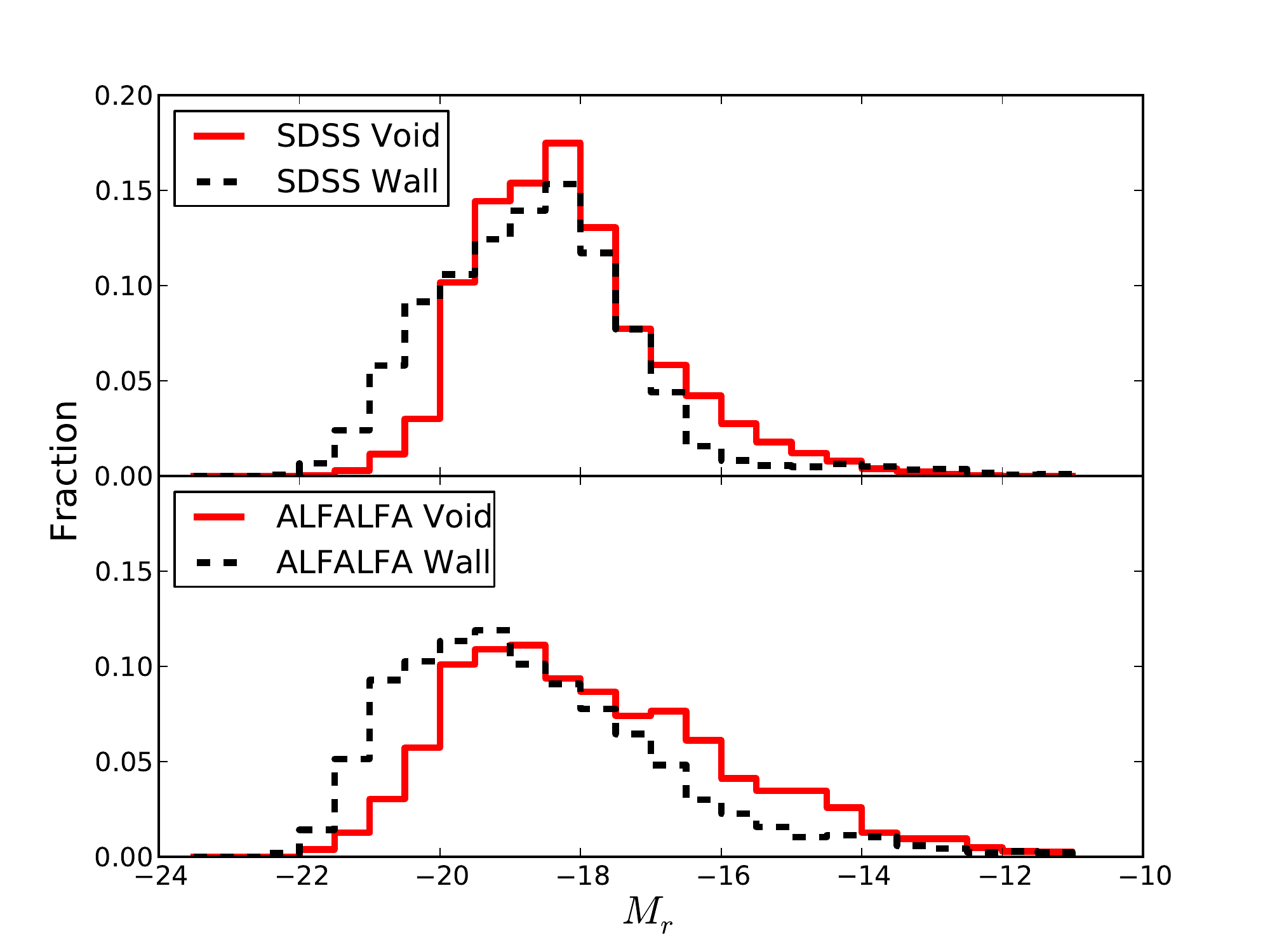}
  \end{center}
  \caption{ Magnitude distribution of \sdssnearby\ (upper) and ALFALFA (lower) 
  void (red) and wall (black) galaxies. 
  The \sdssnearby\ sample contains 7,058 void galaxies 
  and 20,148 wall galaxies. The ALFALFA sample contains 2,611 void detections and 4,566 wall detections. 
  Consistent with previous work (Hoyle et al. 2012), we see that the 
  SDSS void $r$-band magnitude distribution shifts to fainter galaxies. 
  In agreement with this trend, \hi\ selected void galaxies 
  tend to be fainter than their counterparts in denser regions.
  }
  \label{fig:frac_mags}
\end{figure}

Previous studies on the types of galaxies typically found by blind-\hi\ surveys reveal 
late-type, mostly blue galaxies with high star formation rates and 
specific star formation rates \citep{Toribio2009}. 
\cite{Huang2012} provide an in-depth comparison of optically and \hi\ selected samples 
and show that galaxies detected in \hi\ tend to have low star formation efficiencies. 
Here we briefly compare the environmental differences of an \hi\ selected sample 
from the ALFALFA Survey and an optically selected sample 
from the SDSS catalog covering the same volume of sky (\sdssnearby). 

In Figure \ref{fig:frac_color} we compare the optical $u-r$ color distribution of 
\sdssnearby\ (upper) and ALFALFA (lower) galaxies. 
We obtain galaxy colors using SDSS model magnitudes. 
The feature that is most evident is the deficiency of red galaxies in the ALFALFA sample 
compared to the SDSS sample. 
The void and wall SDSS color distributions are bimodal with the second peak appearing 
as a result of the highly populated red sequence, whereas the 
\hi\ color distributions are unimodal with only a slight presence 
of the red sequence in the ALFALFA wall sample. 
While a majority of the ALFALFA detections are blue, we notice the void population on average tends to 
be \textsl{bluer} than the wall population. 
\cite{Hoyle2012} notice a similar trend towards bluer colors in the SDSS void sample. 
In both survey samples, we notice significantly fewer red galaxies in the 
void populations than those in walls. 

In Figure \ref{fig:frac_mags} we present the $r$-band absolute magnitude distributions of the 
\sdssnearby\ (upper) and ALFALFA (lower) samples. 
As shown in \cite{Hoyle2012}, we see a shift toward fainter magnitudes in the SDSS void sample. 
The \hi\ void galaxy sample detects relatively more dwarf galaxies than the optically selected sample. 
This is because the SDSS spectroscopic main galaxy sample is biased against faint, low-surface brightness, patchy galaxies, 
which are more easily detectable in an \hi\ survey. 
Because of the relative abundance of dwarf galaxies in the \hi\ sample we suspect 
the LF of ALFALFA galaxies will have a much steeper slope than the optically selected sample.
We also see that the \hi\ selected wall galaxies are slightly brighter 
than the typical optically selected wall galaxies on average. 
A typical galaxy in a wall will experience gas stripping phenomena (tidal stripping, 
mergers, ram pressure stripping etc), thereby reducing the \hi\ fraction of its baryonic content. 
This reduced \hi\ fraction will contain just enough \hi\ to be detected by a radio survey 
for only the largest/brightest galaxies in the walls. 
We suspect the shift towards brighter galaxies in the mean of the distribution of the ALFALFA wall sample compared to the 
\sdssnearby\ wall sample will result in the ALFALFA wall LF having a somewhat brighter characteristic magnitude than that of \sdssnearby.

As galaxies evolve, they are thought to move from the blue cloud to the 
red sequence after star formation has been quenched. 
Galaxies in voids are typically less evolved--and, therefore, fainter and 
more gaseous--than similar sized galaxies in walls. 
This makes void galaxies easy targets for radio surveys. 
As we will show later in Section \ref{subsubsec:subsets_of_opt_lfs}, 
ALFALFA prefers less evolved galaxies regardless of environment. 

\section{METHOD} 
\label{sec:LF_meth}
\subsection{The SWML Method}
The LF is a measure of the number of galaxies per Mpc$^{3}$ in a 
magnitude range $dM_r$ centered at magnitude $M_r$.
For each measurement of the LF, we find the best fit parameters of 
a \cite{Schechter1976} function of the form 
\begin{eqnarray} 
\Phi(M_{r}) = 0.4 \ln 10 \Phi^* 10^{0.4(M^*-M_r)(\alpha+1)}
\nonumber\\
\times\exp\left(-10^{0.4(M^*-M_r)}\right).
\label{eq:LFSchechter}
 \end{eqnarray} 

We estimate the LF 
of our SDSS and ALFALFA void and wall samples using 
the stepwise maximum likelihood (SWML) method of \cite{EEP1988}. 
For details on how we apply the method, see \cite{Moorman2014} in which we detail the 
two dimensional version of the SWML. 
The SWML method does not retain information about the normalization of the LF; 
therefore, we adjust the amplitude according to the number density of the particular sample of interest. 
For each sample, the wall volume is estimated by taking the difference between the total volume of interest and the volume of the voids within. 
We estimate the best-fit Schechter parameters 
(the normalization factor $\Phi^*$, the characteristic magnitude $M^{*}$, 
and the faint-end slope $\alpha$) to our functions over the 
magnitude range $-22.0<M_{r}<-13.0$ using a least squares estimator. 

\subsection{Errors}
\label{subsec:LF_Errors} 
We estimate errors for each optical LF from three sources.  
The first source is Poisson error which account for $\sim70$ percent of the uncertainties in each bin. 
These errors affect the brightest and faintest ends 
more so than the intermediate magnitude bins, because we have less information in the outermost bins. 
That is, extremely bright galaxies are uncommon in the Universe and 
the faintest galaxies are difficult to detect beyond the very nearby Universe. 

The second source is the error in each bin introduced from the SWML method, described in \cite{EEP1988}. 
These errors account for about 30 percent of the total bin uncertainties. 
As with the Poisson errors, we have less information about the galaxy distribution in the brightest and faintest bins, 
making the uncertainties larger in these bins. 

The third source is an error estimate accounting for the inhomogeneity of large-scale structure, 
using the jackknife method of \cite{Efron1982}. 
For this source of error, we divide our region of interest into 18 subregions, 
and calculate the LF of the galaxy sample excluding one of the 18 regions for each iteration.  
We estimate the variance in each bin after all iterations. 
The jackknife errors account for less than 1 percent of the total error. 
This suggests that the SWML method is relatively robust against large-scale structure.

\section{RESULTS} 
\label{sec:results}
\subsection{LF of Void Galaxies in an Optically Selected Sample} 
\label{subsec:opt_lf} 
\begin{figure}[h]
  \begin{center}
    \includegraphics[scale=0.35,trim= 1.cm 0.5cm 1.25cm 1.75cm, clip=True]{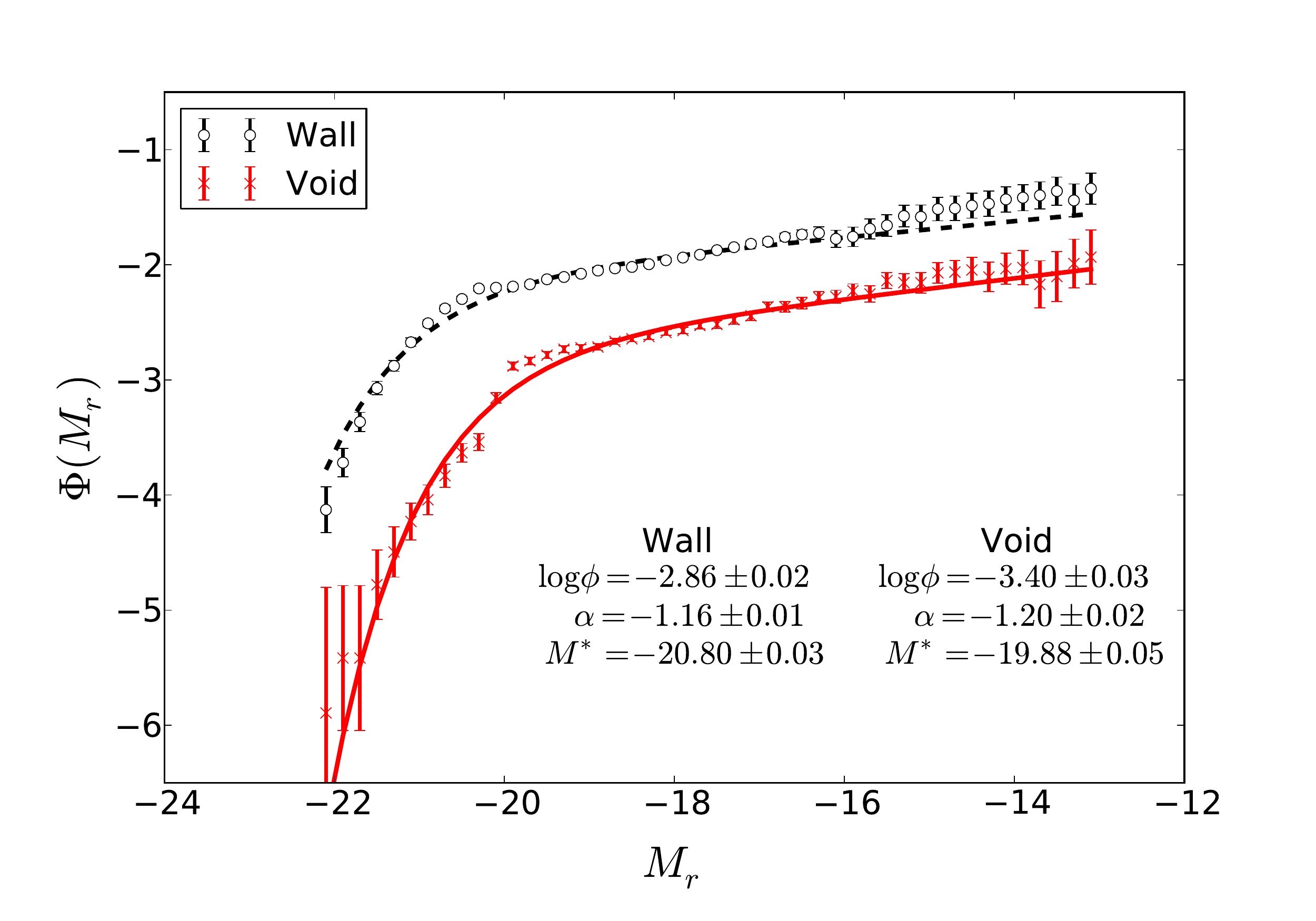}
  \end{center}
  \caption{\small LF of SDSS void (red) and wall (black) galaxies. 
  We find a shift towards fainter galaxies by almost a full magnitude. Consistent with 
  previous studies, we do not find a statistically significant difference in the faint-end slopes.  
    \normalsize}
  \label{fig:void_wall_opt_LF}
\end{figure}
We calculate the LF of our SDSS void and wall galaxies 
down to $M_r\sim-13$ using the methods outlined in Section \ref{sec:LF_meth}. 
In Figure \ref{fig:void_wall_opt_LF}, we present our estimates of the void (red) and wall (black) 
galaxy LFs with the best fitting Schechter functions. The parameters associated with the 
best fit Schechter functions are: $\log\Phi^*= -3.40 \pm 0.03$ $\log($Mpc$^{-3})$, 
$M^*= -19.88\pm 0.05$, and $\alpha= -1.20\pm 0.02$ for void galaxies, 
and $\log\Phi^*=-2.86 \pm 0.02$ $\log($Mpc$^{-3})$, 
$M^*= -20.80\pm 0.03$, and $\alpha= -1.16\pm 0.01$ for wall galaxies. 
An explanation of uncertainties may be found in Section \ref{subsec:LF_Errors}. 
Much like the void/wall LF results found using a previous partial data release of the SDSS \citep{Hoyle2005}, 
we find about a one magnitude shift in $M^*$ towards fainter galaxies in voids. 
It is obvious from Figure \ref{fig:void_wall_opt_LF} that the faint-end slopes of 
both the void and wall LFs are underestimated. For a more accurate estimation of the faint-end slopes, 
we fit a power law to the LF values fainter than $M_r=-18$. These power law slopes 
correspond to a slope of  $\alpha= -1.25\pm 0.02$ for void galaxies and 
$\alpha= -1.27\pm 0.02$ for wall galaxies. 
There is no statistically significant difference between the slopes of the void and wall LFs down to $M_r\sim-13$ 
indicating that there is no relative excess of void dwarfs compared to the wall distribution. 
This faint end slope result is consistent with the predictions of \cite{Cui2011} whose simulations 
show that the faint ends of the LF should remain the same between the most underdense and 
overdense regions. 
This conflicts the predictions of \cite{MathisWhite2002}, although, these authors neglect 
SNe feedback  and background UV radiation 
which should be included when calculating the infall rate of cold gas for low mass haloes. 
The faint end of the void and wall LFs starts to stray from a 
classic Schechter function once we reach the dwarf regime ($M_r>-17$). 
This variation doesn't appear in the analysis of \cite{Tempel2011}, 
because the authors exclude galaxies fainter than $M_r=-17$. 
Excluding these dwarf galaxies from our analysis, we find a faint-end slope that closely matches that of \cite{Tempel2011}.  

We notice a feature at the bright end ($M_r=-20.1$) of both the void and wall LFs:
the void LF drops in amplitude, while the wall LF increases in amplitude. These features 
are an artifact of the void identification process. The void catalog is defined by a volume-limited sample 
corresponding to galaxies with $M_r<-20.1$. By defining the void catalog using galaxies brighter 
than this magnitude, we will see a significant decrease in the number of bright ($M_r<-20.1$) void galaxies 
and an increase in the number of bright wall galaxies.

\subsubsection{Comparing to Previous Observations and Simulations}
\label{subsubsec:prev_LF_obs_sims}
We have measured the galaxy LF in both voids and walls, where the wall environment is effectively 
an average over all higher density regions. 
Comparing the void and wall galaxy LFs reveals an expected shift towards fainter galaxies in voids, 
but does not reveal any dependence of the faint-end slope on large-scale underdensities. 
[Our finding that the void faint-end slope matches the wall faint-end slope 
is similar to the trend found in \cite{Moorman2014} in which we find 
the low-mass slope of the void HIMF closely matches the low-mass slope of the wall HIMF.] 
We suspect that the void faint-end slope closely matches the LF slope of all galaxies in 
denser regions averaged together (wall galaxies) for the following reason. 
\cite{Rojas2004} and \cite{Hoyle2012} show that void galaxies are generally blue, late-type galaxies, 
and \cite{Tempel2011} show that the faint-end slope of late-type galaxies does not vary with environment. 
Thus, one might expect the faint-end slope of the void galaxy LF to be an 
average of the slopes of 
LFs across denser environments where we find both early- and late-type galaxies. 

Comparing our void LF results with previous work on the LFs of galaxy groups and clusters 
reveals a non-monotonic trend in $\alpha$ with environment. 
We find the void regions have a faint-end slope around $\alpha=-1.2$, 
(which is consistent with an earlier result in \cite{Hoyle2005}). 
\cite{Croton2005} find that galaxies in regions with density contrast $-0.5<\rho<0$ 
(isolated galaxies not associated with voids) 
have a flattened faint-end slope of $\alpha\sim-1.0$. 
Additionally, \cite{Tavasoli2015} investigate the effects of local environment within voids 
on the shape of the LF; these authors estimate the LF of $\sim$110 galaxies in 
``sparse'' voids and 111 galaxies in ``populous'' voids brighter than $M_r=-19$ 
making it difficult to truly compare the faint-end slopes with our work. 
Galaxies in denser environments, such as galaxy groups, 
tend to have flat faint-end slopes in the $\alpha\sim-1.0$ to $-1.2$ range, 
where the slopes increase with increasing group mass \citep{Eke2004,Yang2009}. 
The faint-end slope of the composite LFs of galaxy clusters tend to range from $\alpha\sim-1.2$ to $-1.4$ 
\citep{Valotto1997,DePropris2003,Croton2005,McNaught2014}. 
Put together, the overall trend of the faint-end slope with environment is 
average for voids, flattens for ``field'' galaxies, steepens with increasing mass in groups, 
and either remains constant or steepens further when the large-scale density increases to the cluster regime. 

The steepening trend among denser regions is consistent with predictions from the 
hydrodynamic simulations of \cite{Sawala2014}, 
who find that dwarf halos closer to central galaxies typically had higher masses in the earlier Universe, 
making the probability of star formation more likely. That is, galaxies are more likely to form in haloes 
near denser regions than in isolated haloes in the field. 
The steepening of $\alpha$ with increasing density from the field to clusters corroborates this prediction.

\subsection{LF of Void Galaxies in an \hi\ Selected Sample} 
\label{subsec:hi_lf}
As mentioned earlier, the predicted halo mass function has a steep slope of $\alpha=-1.8$, 
whereas the observed LFs (estimated here and in other works) 
have much shallower slopes of $\alpha\sim -1.2$ to $-1.3$. 
\cite{Blanton2005ELL} show that the faint end of the LF can be steepened 
to $\alpha\sim-1.5$ with the inclusion of low luminosity/low surface brightness galaxies. 
To avoid the optical selection bias against low surface brightness galaxies, 
we estimate the optical LF of a sample of \hi\ selected galaxies from the ALFALFA sample. 

\begin{figure}[h]
  \begin{center}
    \includegraphics[scale=0.35,trim= 1.cm 0.5cm 1.25cm 1.75cm, clip=True]{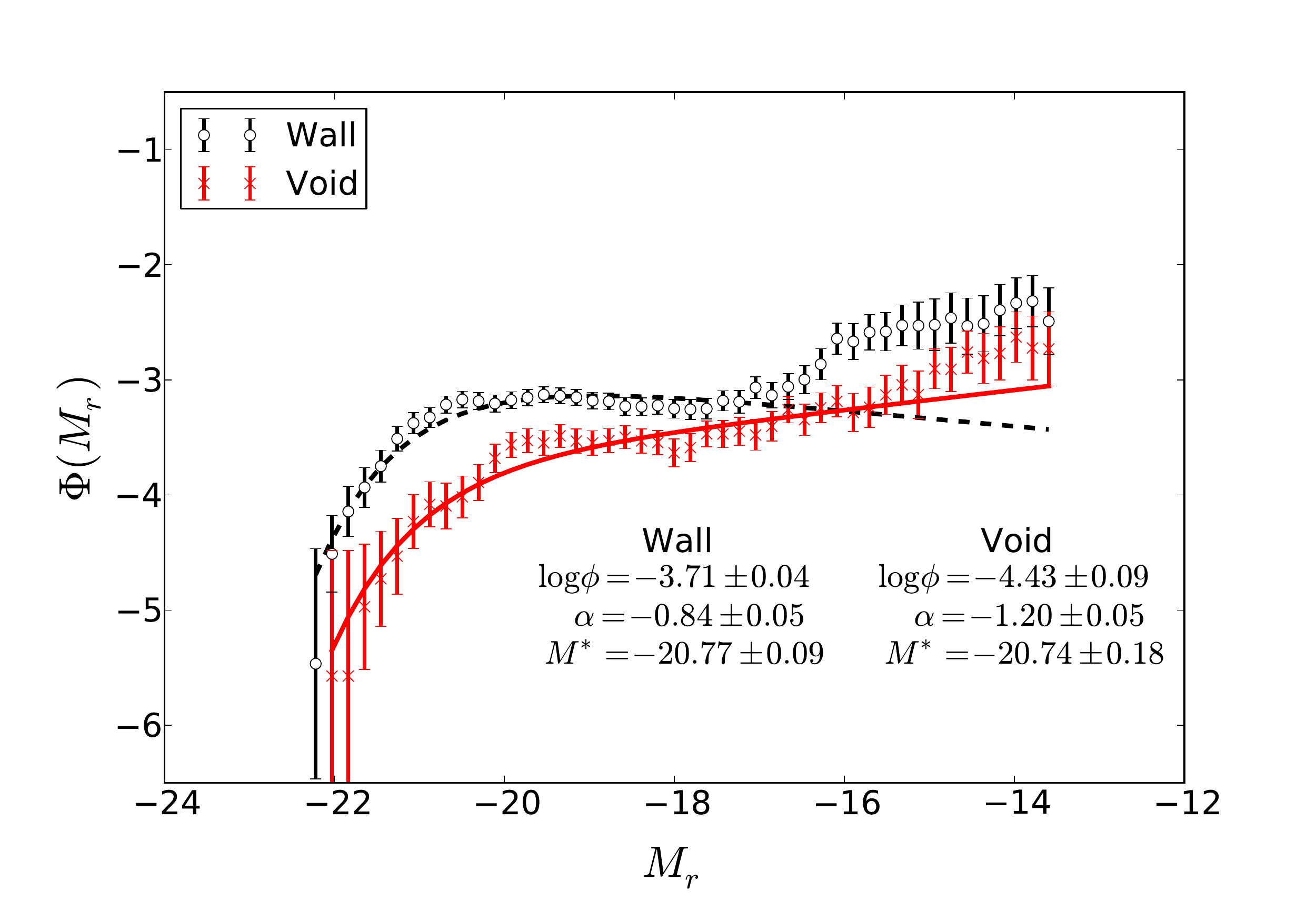} 
  \end{center}
  \caption{\small LF of \hi\ selected void (red) and wall (black) galaxies from $\alpha.40$ Spring Sky. 
  Both void and wall LFs are poorly fit by a simple Schechter function as shown by the curves depicting the best fit 
  Schechter functions found via a least squares estimator.  
    \normalsize}
  \label{fig:void_wall_HI_LF}
\end{figure}
We divide the ALFALFA galaxy sample into void and wall galaxies 
and estimate the $r$-band LF presented in Figure \ref{fig:void_wall_HI_LF}. 
It is clear from the figure that the \hi\ selected LFs are not well fit by a simple Schechter function, 
but for the sake of comparison between the LFs in this work and others, 
we provide the best fitting parameters to a Schechter function found using a least squares estimator. 
For the void sample, we 
estimate the best fitting Schechter parameters 
to be $\log\Phi^*= -4.43 \pm 0.09$ $\log($Mpc$^{-3})$, 
$M^*= -20.74\pm 0.18$, and $\alpha= -1.20\pm 0.05$. 
For the wall galaxy sample, we estimate 
$\log\Phi^*= -3.71\pm 0.04$ $\log($Mpc$^{-3})$, 
$M^*= -20.77\pm 0.09$, and $\alpha= -0.84\pm 0.05$.
See Section \ref{subsec:LF_Errors} for an explanation of uncertainties. 
The curves in Figure \ref{fig:void_wall_HI_LF} show the Schechter functions associated with 
these best-fit parameters; however, the Schechter fits underestimate both the bright and faint ends of the ALFALFA LFs. 

\begin{figure}[h]
  \begin{center}
    \includegraphics[scale=0.475,trim= 1.5cm 0.750cm 1cm 1.5cm, clip=True]{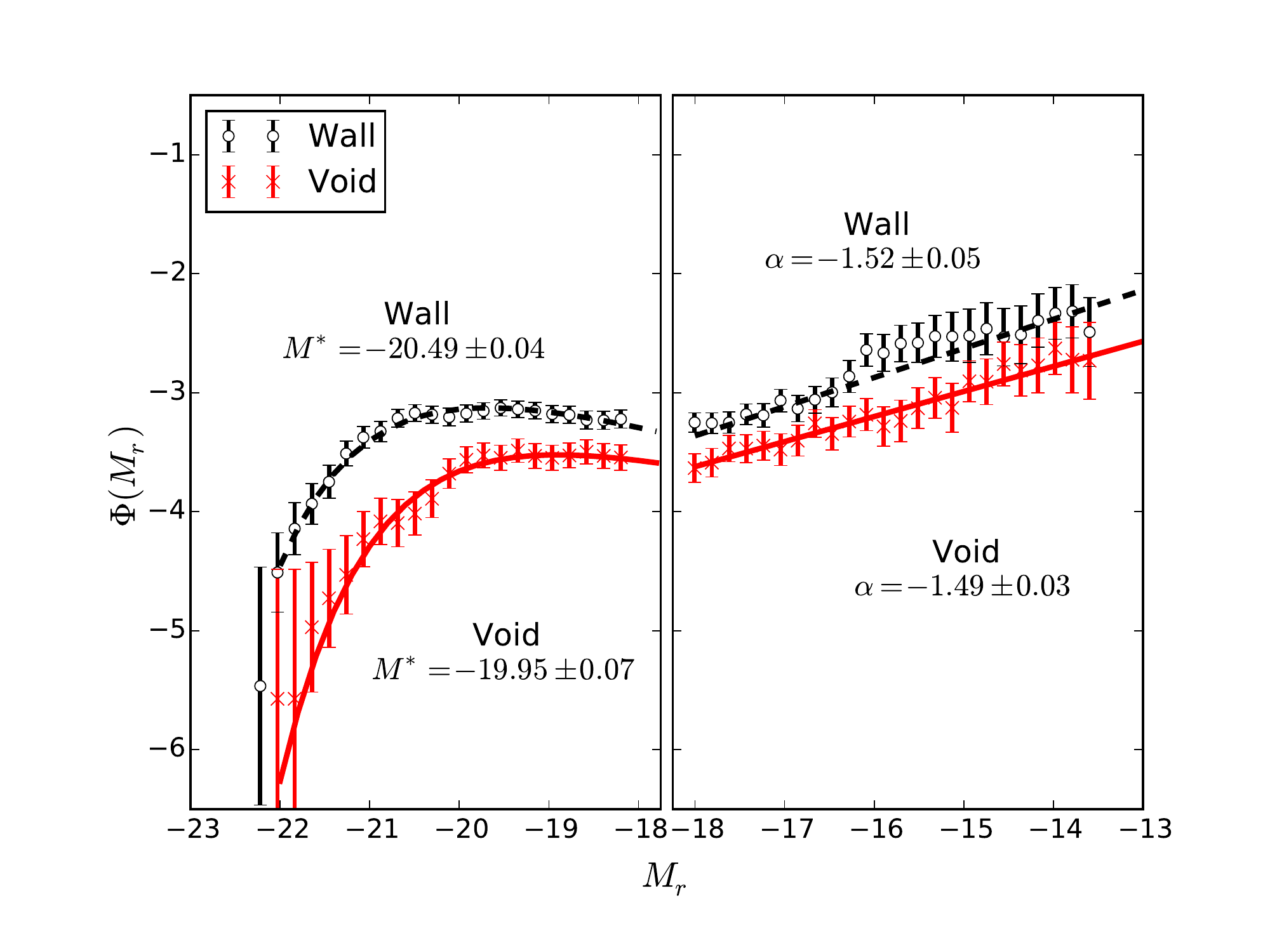} 
  \end{center}
  \caption{\small Separately fit LF parameters of the brightest and faintest \hi\ selected 
  void (red) and wall (black) galaxies from $\alpha.40$ Spring Sky. 
  We identify the $M^*$ parameter by fitting a Schechter function only to the bright 
  ($M_r<-18$) end of the LF to reveal the true shift in the 
  characteristic magnitude between voids and walls. The curves to the left of $M_r=-18$ represent 
  the Schechter fits to the bright galaxy distribution. 
  We separately fit a power law to the faint end of the LF. The lines plotted at the faint end represent 
  the best fit slopes.  
    \normalsize}
  \label{fig:bright_opt_LF}
\end{figure}
At the bright end, the best-fit Schechter function predicts the characteristic magnitudes of 
both the void and wall samples to be too bright. 
To get a realistic sense of the shift in $M^*$ between voids and walls,
we fit only the bright end ($-22\le M_r\le-18$) of the LFs revealing a shift in $M^*$ towards fainter magnitudes 
in voids by about one half 
of a magnitude (see the left panel of Figure \ref{fig:bright_opt_LF}). The direction of this shift is consistent with 
the shift in the dark matter halo mass function of extended Press-Schechter theory \citep{Goldberg2005}. 
The magnitude of the shift is 
significantly less than the shift in the predicted halo mass function as well as the shift in $M^*$ 
of the SDSS LF found in the previous section. 
The reason for this smaller shift is that typically the brightest galaxies in denser regions 
have burnt through their cool gas, lost their gas during mergers, and/or 
had it stripped away via tidal stripping, ram pressure stripping, etc. 
Therefore, a survey looking for cool neutral gas will detect the brightest galaxies 
within denser regions far less frequently than an optically selected sample. 
While an optically selected spectroscopic sample may be biased against faint, 
low-surface brightness galaxies, an \hi\ selected sample is biased against massive, red galaxies. 

At the faint end, the best fitting Schechter function also severely underestimates 
the faint end slopes of the void and wall LFs, estimating a much shallower slope than actually observed. 
To more accurately determine the faint end slopes, 
we fit a power law function to only the faint ends ($-18<M_r\le-13$) of the LFs. 
The new faint end slopes of the \hi\ selected LFs 
(shown in the right panel of Figure \ref{fig:bright_opt_LF}) are 
$\alpha=-1.49\pm0.03$ for voids and $\alpha=-1.52\pm0.05$ for walls. 
We see no statistically significant difference in the 
faint-end slopes of the void and wall LFs. 
The faint-end slopes of the ALFALFA void and wall LFs appear to be independent of 
large-scale environment. We suspect this is largely due to the 
relatively large number of late-type galaxies (which generally dominate the faint-end of the LF) 
present in the ALFALFA sample compared to massive, elliptical galaxies. 
\cite{Tempel2011} find the LF of spiral galaxies is independent of environment, 
while the faint-end slope of the observed elliptical galaxy LF 
steepens with increasing large-scale densities. 

To more directly compare the void and wall distributions of the \hi\ and optically selected samples, 
we need to compute the optical LF of SDSS and ALFALFA galaxies over the same volume of sky. 
Therefore, we measure the LF of the \sdssnearby\ sample and estimate its characteristic magnitude and 
faint-end slope in the following way. 
For each sample (SDSS, \sdssnearby, and ALFALFA), we split each LF into a bright and faint end, divided at $M_{r}=-18$. 
The bright end is fit with a Schechter function, from which we obtain the sample's characteristic magnitude, $M^*$. 
Each faint end is separately fit with a power law, from which we obtain the faint end slope of each LF. 
We provide these fits to the data in Table \ref{tab:fits}. 
Note the $M^*$ and $\alpha$ parameters in the table are not the same 
related parameters estimated in equation (\ref{eq:LFSchechter}). 
\begin{deluxetable}{lccc} 
	\tablewidth{0pt}
	\tablecaption{Separately-fit bright and faint ends\label{tab:fits}}
	\tablehead{\colhead{Sample} & \colhead{LSS} & \colhead{$M_{r}^*$} & \colhead{$\alpha$}} 
	\startdata
	SDSS 	& void 	& -19.32$\pm$0.03 	& -1.25$\pm$0.02 	\\ 		
	SDSS 	& wall 	& -20.54$\pm$0.02 	& -1.27$\pm$0.02 	\\ \hline	
	\sdssnearby 	& void 	& -19.30$\pm$0.06 	& -1.31$\pm$0.04 	\\		
	\sdssnearby & wall 	& -20.55$\pm$0.03 	& -1.23$\pm$0.03 	\\  \hline	
	ALFALFA 	& void 	& -19.95$\pm$0.07 	& -1.49$\pm$0.03 	\\		
	ALFALFA 	& wall 	& -20.49$\pm$0.04 	& -1.52$\pm$0.05 	\\  \hline	
	\enddata
	\tablecomments{Best fit power law and Schechter function parameters 
	to the optical LFs of the SDSS and ALFALFA samples across environments. 
	Each LF was split into a bright end and faint end, separated at $M_{\rm r}=-18$. 
	Each bright end was fit by a Schechter function; from this function, 
	we extract the characteristic magnitude parameter, $M_r^*$. 
	Each faint end ($M_r>-18$) was fit by a power law function. 
	From this fit, we extract the faint end slope parameter, $\alpha$.}
\end{deluxetable}

Two interesting comparisons arise from this table. 
The first comes from comparing 
the full SDSS and \sdssnearby\ fits, which 
gives us information on how the selected volume affects the shape of the LF.  
We find it interesting that reducing the area and redshift affects the void faint end slope. 
The faint end slopes of the full SDSS sample are more accurate, 
because we are averaging over more structure. 
The large-scale structure within the nearby volume may be atypical of the full Universe. 
The second comparison worth making is between 
the ALFALFA and \sdssnearby\ fits, which 
gives us information on how sample selection affects the shape of the LF. 
We see that using an \hi\ selected sample produces a much steeper faint end slope than 
an optically selected sample. 
This is due mostly to the inclusion of extremely low luminosity 
galaxies in \hi\ surveys. 
The SDSS photometry affects the main galaxy sample 
target selection. The selection process is biased against low-surface brightness galaxies, 
so these faint galaxies are excluded from our optical sample. 
(Refer to Figure \ref{fig:frac_mags} for a comparison of the $M_r$ distributions of the two samples.) 
The power law fits to the faint-end slopes match closely those shown in \cite{Blanton2005ELL} who 
investigate the effects of extremely low luminosity galaxies on the faint-end slope of the galaxy LF. 
Again, these authors find that the slopes increase from $\alpha=-1.3$ to $\alpha=-1.5$ when 
adjusting for the incompleteness of the SDSS spectroscopic sample at low surface brightnesses. 

\subsection{Comparing the Optically and \hi\ Selected LFs} 
\label{subsec:LF_compare_hi_and_opt}  
\label{sec:LF_results} 
\begin{figure}[h]
  \begin{center}
    \includegraphics[scale=0.35,trim= 1.cm 0.5cm 1.25cm 1.75cm, clip=True]{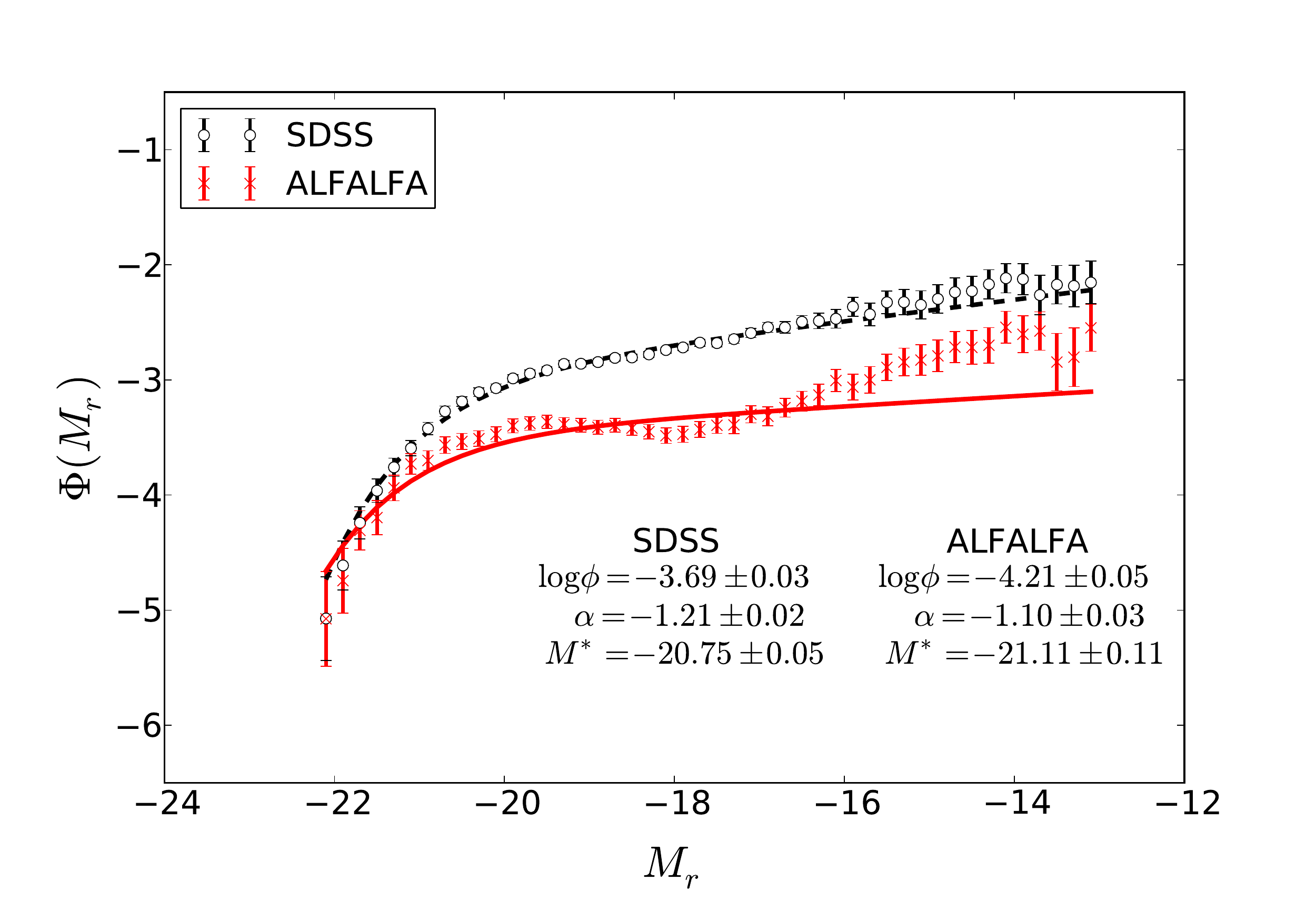}
  \end{center}
  \caption{\small LF of optically selected galaxies from \sdssnearby\ with the best-fit 
  Schechter function (black) and the LF of \hi\ selected galaxies from the $\alpha.40$ Spring Sky 
  with the best-fit Schechter function (red). 
    \normalsize}
  \label{fig:opt_vs_hi_LF}
\end{figure}
In Figure \ref{fig:opt_vs_hi_LF}, 
we present the global LF of the Spring Sky subsample of the $\alpha.40$ data set 
as well as that of \sdssnearby\ sample. 
We see in this figure, as well as the previous subsection, that the optical LF of 
\hi\ selected galaxies has a clearly different shape than that of the optically selected galaxies. 
The optically selected LF is reasonably well fit by a Schechter function with estimated 
parameters $\log\Phi^*= -3.69 \pm 0.03$ $\log($Mpc$^{-3})$, 
$M^*= -20.75\pm 0.05$, and $\alpha= -1.21\pm 0.02$. 
It is clear from the figure that the \hi\ selected global LF is 
not fit well by a Schechter function. 
The best fit Schechter parameters are 
$\log\Phi^*= -4.21 \pm 0.05$ $\log($Mpc$^{-3})$, 
$M^*= -21.11\pm 0.11$, and $\alpha= -1.10\pm 0.03$. 
One of the most notable aspects of the \hi\ selected LF is a broad, dip-like feature around $M_r=-18$ 
followed by a sharp upturn at fainter magnitudes. 
Evidence of a similar dip and upturn is seen in \cite{Zwaan2001} in the LF of \hi\ selected galaxies from the 
Arecibo \hi\ Strip Survey, though the authors only use a sample of 60 \hi\ sources and attribute 
the features of the faint-end slope to low number statistics. 
For the first time, we show statistically significant evidence for a population of LSB dwarf galaxies present in the 
\hi\ selected optical galaxy LF. 

We suspect that the wide dip present in the ALFALFA LF is 
the result of a linear combination of different types of optically selected galaxies, 
i.e. dwarfy-starbursting galaxies at the faint end and gas-rich spirals at the bright end. 
In the next section, we will investigate different combinations of optically selected galaxies 
that may produce similar features and further split the \hi\ sample to see which properties of the 
galaxies may be causing these features. 
Additionally, we wish to note that \cite{Papastergis2012} see a similar, albeit much shallower, dip feature 
in the baryonic mass function of ALFALFA galaxies. While these two phenomena may be related, 
it does not explain the severity of the dip seen in the ALFALFA galaxy LF.   

\begin{figure}[h]
  \begin{center}
    \includegraphics[scale=0.39,trim = 1.5cm 0.9cm 1.95cm 2cm, clip=True]{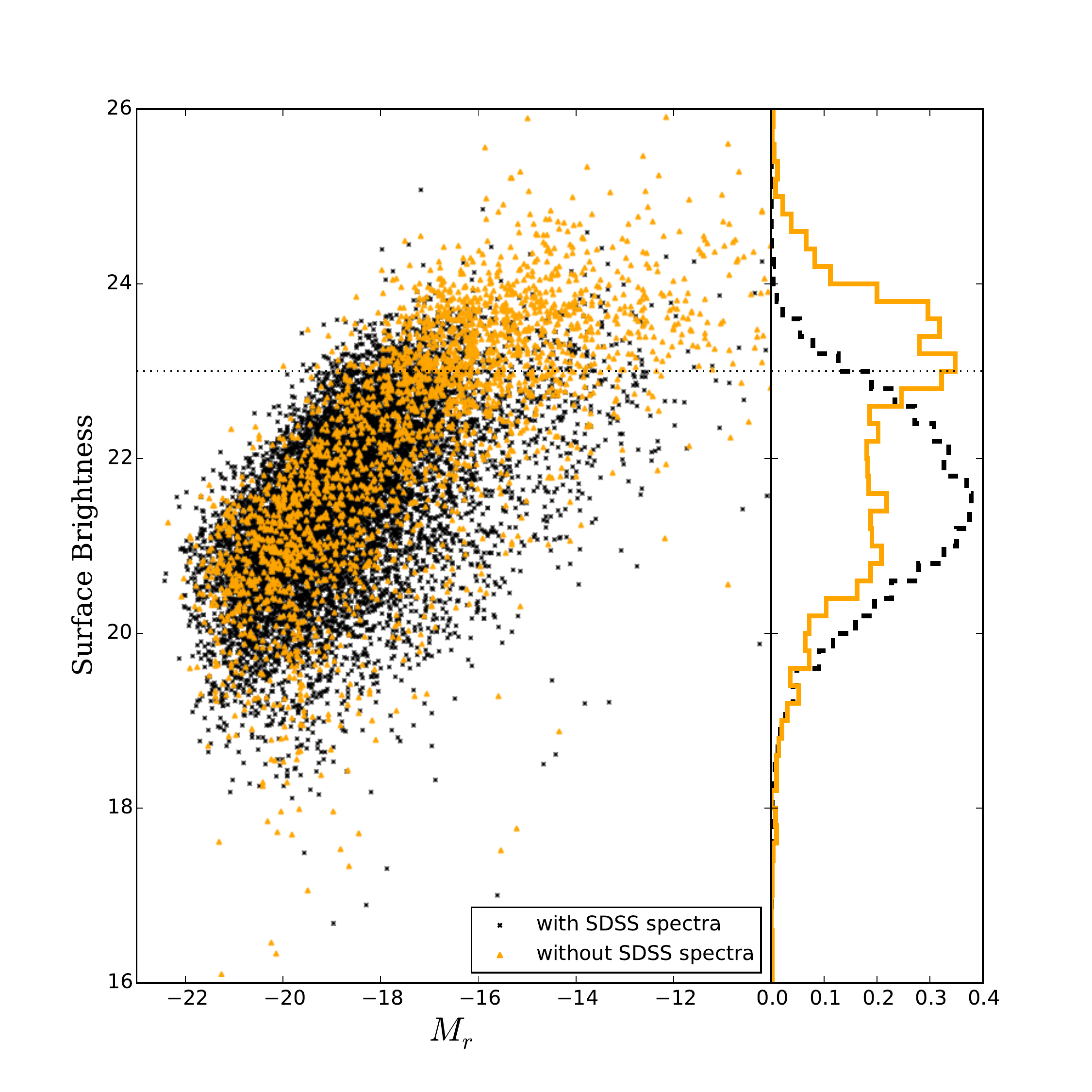}
  \end{center}
  \caption{Distribution of ALFALFA galaxies with (black) and without (orange) SDSS spectra in SB--$M_r$ space. 
  ALFALFA contains a significant population of LSB dwarf galaxies not present in the SDSS main galaxy spectroscopic survey.  
  The histogram (right panel) depicts the normalized SB distribution of ALFALFA galaxies  
  with (dashed black) and without (orange) SDSS spectra.
  The dotted line denotes the LSB limit of \cite{Blanton2005ELL}, 
  in which galaxies with $\text{SB}>23$ are considered to be LSB detections.} 
  \label{fig:LSB_hist}
\end{figure}

Following the previous subsection, we fit 
the bright and faint ends separately with a Schechter function and power law, respectively. 
Again, the bright/faint division takes place at $M_r=-18$, where we see the dip-like feature. 
The characteristic magnitude of the \hi\ selected sample brighter than $M_r=-18$ is $M^*=  -20.48\pm 0.04$. 
The faint end slope, estimated via a power law fit to the LF fainter than $M_r=-18$, is $\alpha=-1.47 \pm 0.02$. 
As mentioned above, the dramatic steepening of the faint end slope of the \hi\ selected LF is 
most likely due to the inclusion of extremely low luminosity/LSB galaxies \citep{Blanton2005ELL}. 
Figure \ref{fig:LSB_hist} shows the presence of a population of LSB galaxies in our ALFALFA sample 
that do not appear in the spectroscopic SDSS sample. 
The SDSS spectroscopic sample is biased against these very faint galaxies, 
thus we do not see such a drastic upturn of the optically selected LF. 
\cite{Zwaan2001} find that gaseous LSB galaxies make up only 5$\pm$2 per cent of the luminosity density, 
suggesting that there shouldn't be low surface brightness effects from an \hi\ survey, 
but these galaxies dominate the shape of the LF at the very faint end steepening the slope by $\Delta\alpha\sim0.26$. 

\subsubsection{Subsets of the Optically Selected LF}
\label{subsubsec:subsets_of_opt_lfs}
\begin{figure}[h]
  \begin{center}
    \includegraphics[scale=0.35,trim= 1.cm 0.5cm 1.25cm 1.75cm, clip=True]{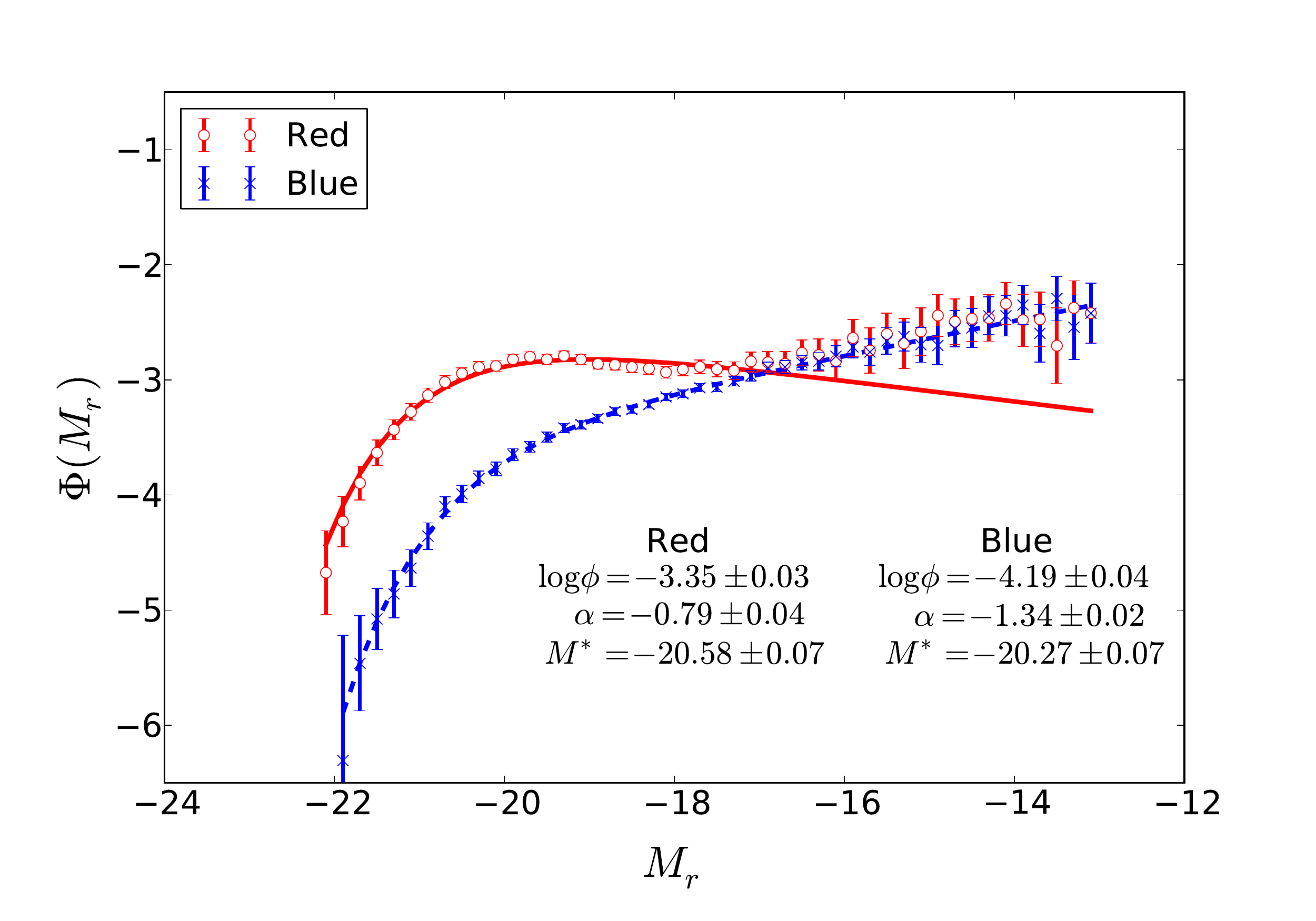}
  \end{center}
  \caption{ LF of red and blue optically selected galaxies from SDSS DR7 with best-fit 
  Schechter functions. The red galaxy LF shows similar features to that of the \hi\ sample.
   }
  \label{fig:opt_color_lfs}
\end{figure}

Given the dramatic differences in the overall shapes of the ALFALFA and \sdssnearby\ LFs, 
we would like to determine if the unique shape of the ALFALFA LF is reproducible using a combination 
of populations from the \sdssnearby\ sample. 
Knowing that most galaxies in an \hi\ survey tend to be blue, late-type galaxies, 
we first split the \sdssnearby\ sample into blue and red galaxies using the color cut from \cite{Moorman2014}: 
$u-r=-0.09M_r+0.46$, where $M_r$ is an object's $r$-band absolute magnitude. 
We consider a galaxy with $u-r$ color less than the value given by the equation to be blue, 
otherwise the galaxy is considered red. 
Our resulting subsamples contain 16,548 blue galaxies and 11,147 red galaxies.
We estimate the LF of blue and red samples, and, surprisingly, find that the red galaxy LF produces a dip 
similar to, albeit much shallower than, that of the ALFALFA sample. 
See Figure \ref{fig:opt_color_lfs} for the LFs of optically selected blue and red galaxies. 

\begin{figure}[h]
  \begin{center}
    \includegraphics[scale=0.435,trim= 0.85cm 0.3cm 1.25cm 0.75cm, clip=True]{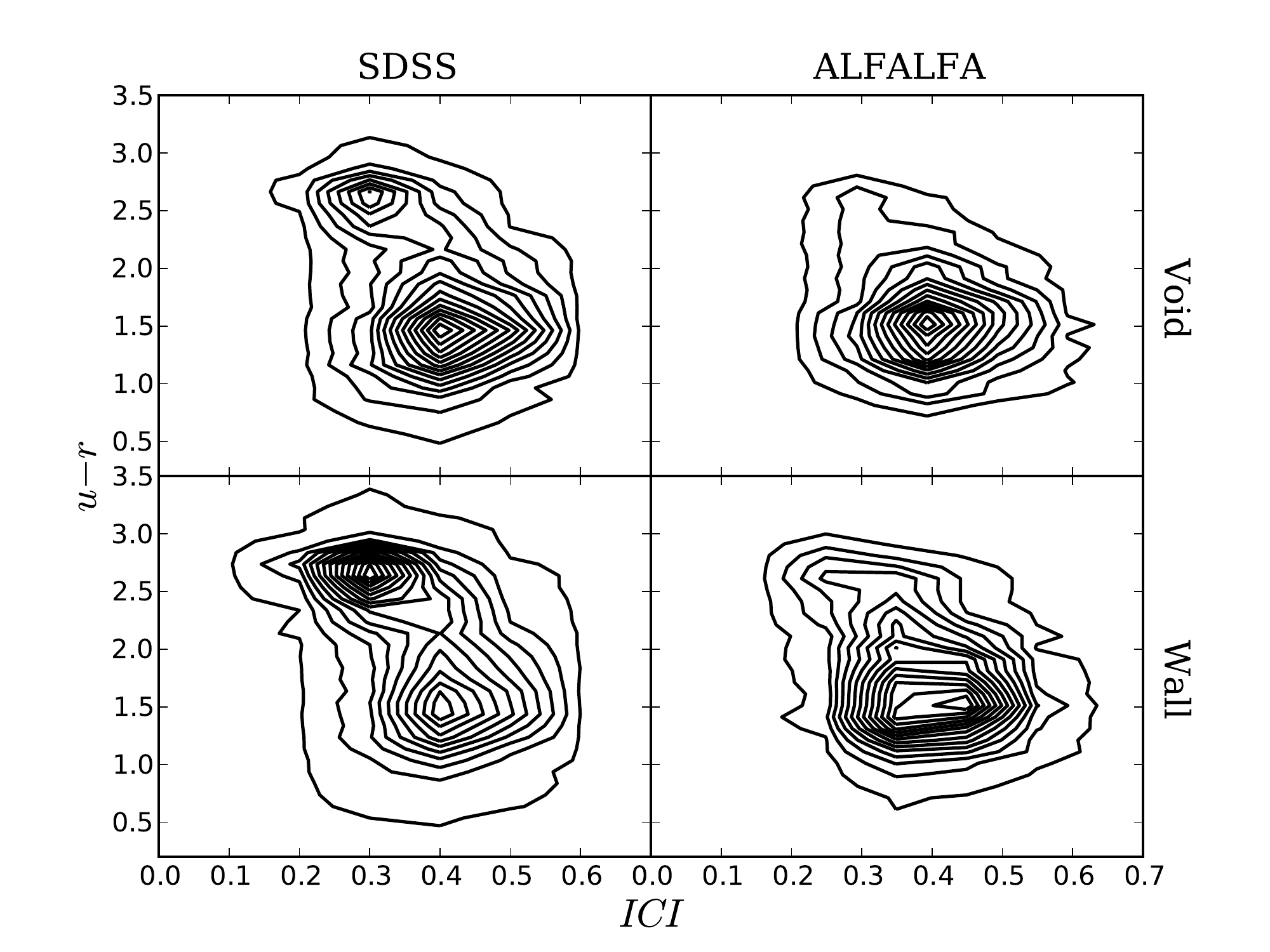}
  \end{center}
  \caption{Normalized inverse concentration index vs. color density contours of void (top row) and wall (bottom row) 
  galaxies in \sdssnearby\ (left column) and ALFALFA (right rolumn). 
  Both void and wall samples in SDSS contain both late- and early-type galaxies. 
  The void galaxies are preferentially later-type, while the wall galaxies are primarily more evolved. 
  Void and wall detections in ALFALFA both prefer blue, late-type galaxies. 
    }
  \label{fig:ici_vs_color_figs}
\end{figure}

Suspecting that the red sample is composed of both large elliptical galaxies 
as well as edge-on spirals reddened by dust, we split the \sdssnearby\ red sample into two categories of 
morphological type. 
We make morphological cuts based on a galaxy's inverse concentration index (ICI), 
which is shown to be correlated with morphological type 
\citep{Shimasaku2001}.  
The ICI is defined to be $c_{in}=R_{50}/R_{90}$, where $R_{50}$ and $R_{90}$ 
are the radii containing 50\% and 90\% of the integrated Petrosian flux of a galaxy.
In Figure \ref{fig:ici_vs_color_figs}  
we present the normalized distribution of each sample's color vs. morphological type 
via the galaxies' inverse concentration indices. 
It is clear from the figure that ALFALFA tends to detect less evolved 
galaxies than SDSS, regardless of environment. 
Within the \sdssnearby\ sample, it appears that the void galaxies span the range of 
inverse concentration indices of ICI=0.2-0.6, whereas the 
wall galaxies are primarily early-type galaxies (ICI$<$0.42). 

\begin{figure}[h]
  \begin{center}
    \includegraphics[scale=0.35,trim= 1.cm 0.5cm 1.25cm 1.75cm, clip=True]{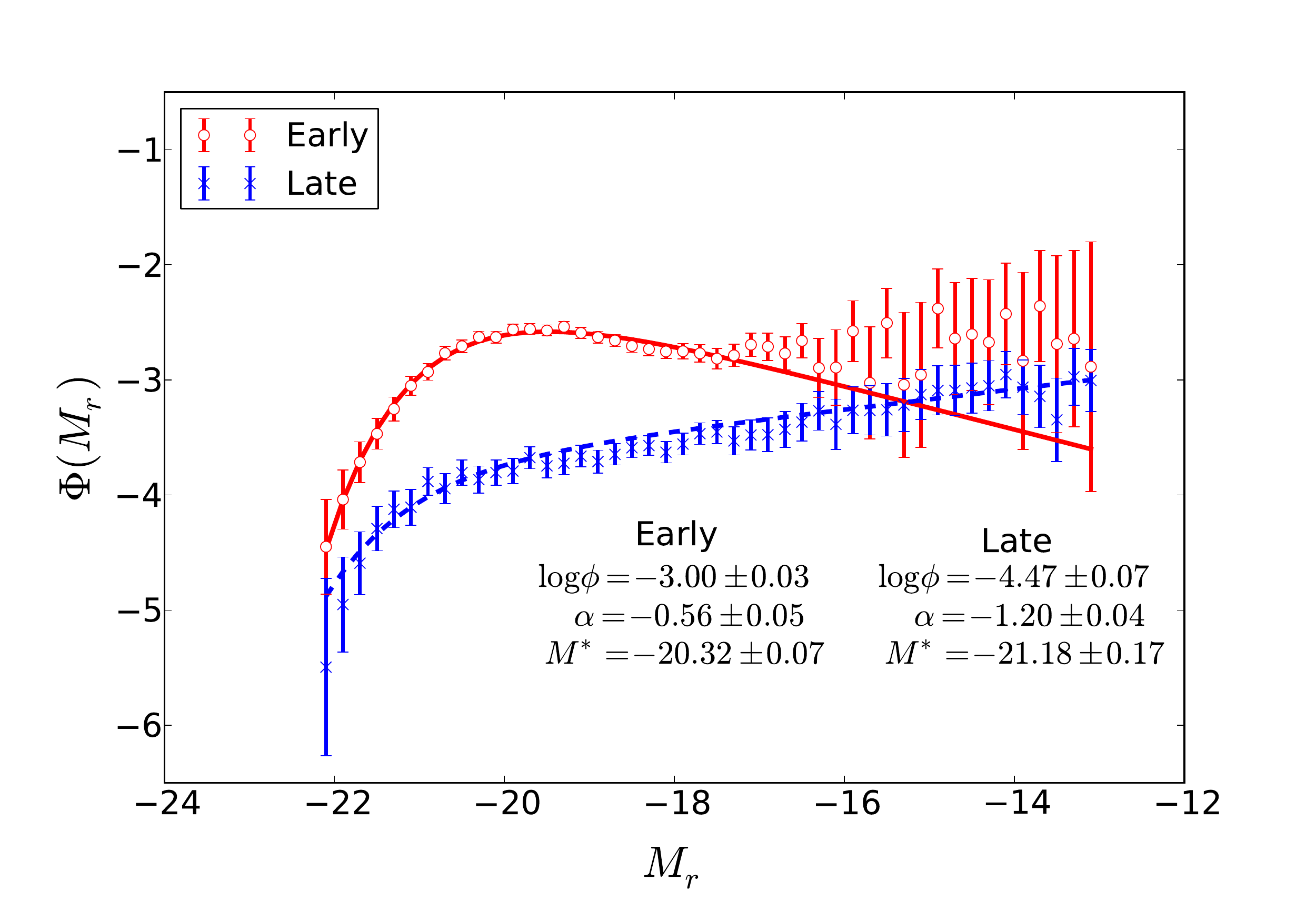}
  \end{center}
  \caption{\small LF of red optically selected galaxies from \sdssnearby\ with best-fit 
  Schechter functions split by morphology, as determined by the inverse concentration index. 
  The early-type population has a relatively flat distribution and consists 
  primarily of large elliptical galaxies. 
  The late-type population has distribution that is well fit by a Schechter function 
  with a faint-end slope similar to that of the global sample. 
    \normalsize}
  \label{fig:red_early_vs_late_types}
\end{figure}
\begin{figure}[h]
  \begin{center}
    \includegraphics[scale=0.35,trim= 1.cm 0.5cm 1.25cm 1.75cm, clip=True]{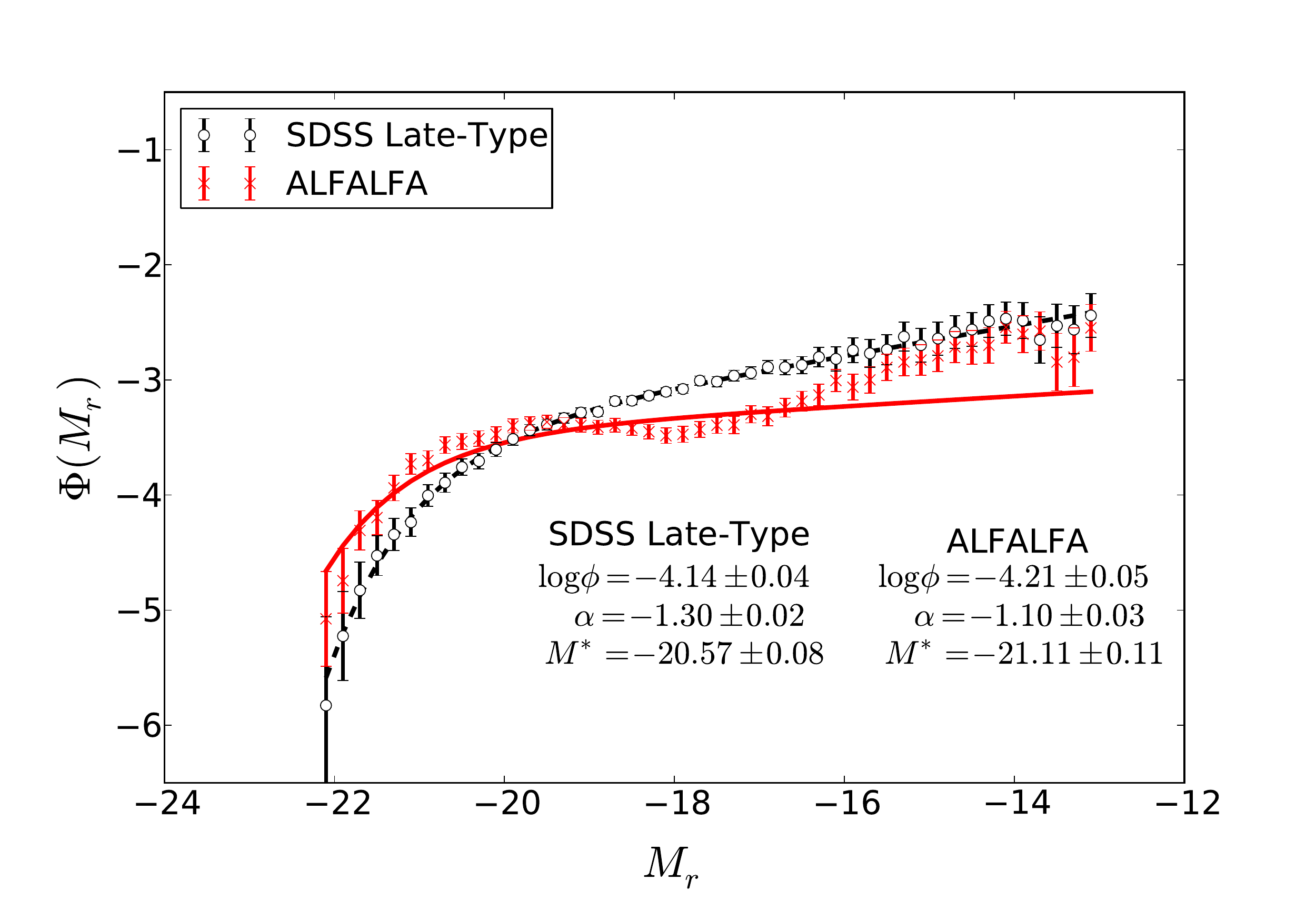}
  \end{center}
  \caption{\small LFs of ALFALFA galaxies with best-fit 
  Schechter functions (red) and \sdssnearby\ late-type galaxies with best fit Schechter functions (black). 
  Given that \hi\ surveys tend to observe late-type galaxies, we would expect the ALFALFA LF 
  to resemble the late-type \sdssnearby\ LF. Surprisingly, this is not the case. 
    \normalsize}
  \label{fig:hi_vs_spiral_lf}
\end{figure}

From the \sdssnearby\ sample, we have 8127 red elliptical galaxies and 3020 reddened spiral galaxies.
As expected, we see, in Figure \ref{fig:red_early_vs_late_types}, 
that splitting the red sample by ICI produces two distinct functions. 
The early-type galaxies have a relatively flat distribution with a 
small dip-like feature present around $M_r=-18$; 
while the late-type galaxies are well fit by a Schechter function with a faint-end slope similar to that of the global SDSS sample. 
The red, late-type galaxy LF has a similar characteristic magnitude 
to the blue galaxy LF, but has a shallower slope. 
\cite{Zwaan2001} show that their \hi\ selected LF closely matches that of late-type 
galaxies found in optical surveys. We find that this is not the case for the ALFALFA LF, 
because of the significant population of massive gas-rich galaxies mentioned in \cite{Huang2012}. 
See Figure \ref{fig:hi_vs_spiral_lf} comparing the ALFALFA LF and the \sdssnearby\ late-type galaxy LF. 

\subsubsection{Subsets of the \hi\ Selected LF}
\label{subsubsec:subsets_of_hi_lfs}
\begin{figure}[h]
  \begin{center}
    \includegraphics[scale=0.35,trim= 1.cm 0.5cm 1.25cm 1.75cm, clip=True]{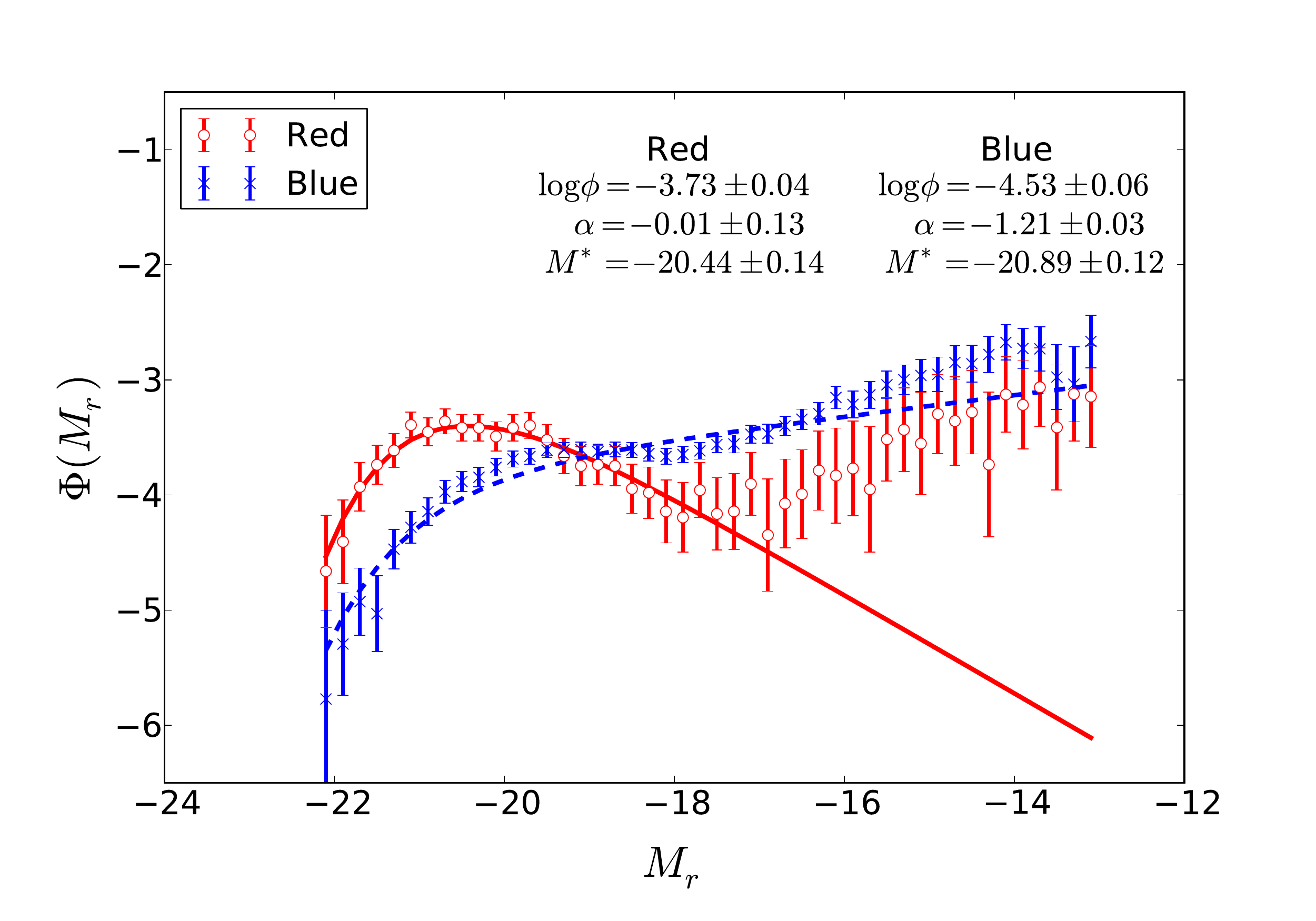}
  \end{center}
  \caption{\small LF of red and blue \hi\ selected galaxies from ALFALFA with best-fit 
  Schechter functions. Both distributions feature a dip around $M_r=-18$. 
  The red galaxy distribution is relatively flat and is dominated 
  by large bright galaxies. The blue distribution has a steeper faint-end slope and is 
  most likely dominated by dwarfy star-bursting galaxies. 
    \normalsize}
  \label{fig:hi_color_lfs}
\end{figure}
No known \hi\ selection effects could be responsible for producing such a wide dip in the galaxy LF. 
In attempts to directly determine what populations of galaxies 
could be influencing the shape of the \hi\ selected optical LF, 
we divide the ALFALFA sample by color into red (1376) and blue (6234) galaxies 
and compute the LFs. 
As shown in Figure \ref{fig:hi_color_lfs}, the dip in the LF remains present 
in the \hi\ selected red galaxy sample, and we see a much less prevalent inflection 
at the same magnitude for the \hi\ selected blue galaxy sample. 
Splitting the red and blue distributions by morphological type, based on ICI, 
does not reveal the origin of the dip feature present in the ALFALFA LF. That is, 
no subsample of ALFALFA galaxies removes the dip feature from the LF. 

\section{CONCLUSIONS} 
\label{sec:LF_conc}
Using the SDSS DR7 void catalog obtained from \cite{Pan2012}, 
the KIAS-VAGC SDSS DR7 galaxy catalog from \cite{Choi2010}, and the
ALFALFA $\alpha.40$ catalog from \cite{Haynes2011}, we measure the optical LF of 
75,063 optically selected void galaxies with $r$-band magnitudes ranging from $-22.0<M_r<-13.0$ and 
2,611 \hi\ selected void galaxies with $r$-band magnitudes ranging from $-22.0<M_r<-13.0$. 
We find that sample selection plays a large role in determining the shape of the LF. 
Within a given data set, the large-scale environment affects the 
value of the characteristic magnitude of the LF, in that the characteristic magnitude shifts toward 
fainter values in cosmic voids. 
The environmental effects on the faint end slope vary with the volume over which we look. 

We find that the LF of void galaxies from the full SDSS sample is well fit by a Schechter
function with parameters 
$\log\Phi^*= -3.40 \pm 0.03$ $\log($Mpc$^{-3})$, 
$M^*= -19.88\pm 0.05$, and $\alpha= -1.20\pm 0.02$.
For galaxies residing in higher density regions, we find the best fit Schechter parameters
to be  $\log\Phi^*=-2.86 \pm 0.02$ $\log($Mpc$^{-3})$, 
$M^*= -20.80\pm 0.03$, and $\alpha= -1.16\pm 0.01$. 
Our findings suggest that the location of the LF is dependent on environment. 
That is, the characteristic magnitude, $M^*$, shifts to fainter magnitudes in voids. 
For the optically selected sample, the shift in $M^*$ is about one magnitude, 
consistent with an earlier partial SDSS data release \citep{Hoyle2005}. 
The direction of the shift is consistent with extended Press-Schechter theory  
which states that the dark matter mass function should shift to lower masses in underdense regions \citep{Goldberg2005}. 
When we fit Schechter functions to the void and wall LFs over the range $-22.0<M_r<-13.0$, 
we see a small environmental dependence on the faint end slope Schechter function parameter. 
However, the best fitting Schechter function underestimates the faint end slope 
of the void and wall optical LFs. 
To account for this, we fit a power law to only the faint ends of the 
void and wall LFs (refer back to Table \ref{tab:fits}). 
We find that the true faint end slope of the optically selected void galaxy LF 
is the same as that of the wall galaxy LF. 
It is important to note that the faint-end slope of the LF varies among isolated galaxies, groups, and clusters. 
The faint-end slope of the void LF matches the faint-end slope of all dense regions averaged together. 

Limiting the SDSS sample to the volume over which $a.40$ and SDSS overlap (\sdssnearby), 
yields similar results to the full SDSS sample, regarding $M^*$. 
That is, the characteristic magnitude in voids shifts fainter by about a full magnitude. 
However, we do see an environmental dependence on the faint end slope 
when we fit a power law to the void and wall faint end slopes. Over this reduced volume, 
the faint-end slope of void galaxies, estimated by a power law fit to the faint end, 
steepens to $\alpha=-1.31\pm0.04$ as opposed to $\alpha=-1.23\pm0.03$ in walls. 
The difference in faint-end slopes in the \sdssnearby\ sample is due to the fact that we 
are not averaging over as much structure as in the full SDSS sample. 
The large-scale structure within the nearby volume 
is likely not representative of the full volume of the local Universe; thus, 
we suspect the full SDSS LF estimates are better representatives of the 
actual void and wall LFs of the local Universe. 
The estimated faint-end slopes of galaxy LFs are highly dependent on the 
volume over which we observe, and 
we provide these results solely to compare the effects of sample selection on the optical LF. 

The optical LF of ALFALFA galaxies within the $\alpha.40$ Spring Sky region 
has a very wide, dip-like feature around $M_r=-18$; thus, the ALFALFA optical LF 
is not well fit by a Schechter function. 
We are currently unsure of the origin of the dip-like feature in the ALFALFA LF, but we 
suspect there may be a connection between 
the inclusion of massive, gas-rich galaxies in the ALFALFA sample and 
the bizarre shape.  
We do point out, however, that this feature cannot 
be explained by a simple combination of populations of optically selected galaxies. 
Splitting the \hi\ sample by color, via SDSS model magnitudes, 
and morphological type, via the inverse concentration index, 
reveals a significant population of late-type reddish galaxies. 
While the red elliptical galaxy LF from the \sdssnearby\ sample produced a relatively small dip 
around $M_r=-18$, the red elliptical population in ALFALFA is small compared to other galaxy types. 
Additionally, the dip feature is also seen in the ALFALFA blue galaxy LF. 
Thus, it is improbable that the red elliptical distribution alone caused the shape. 
The magnitude range over which we find this feature corresponds to 
a stellar mass range of $\log(M_*/M_{\sun})\sim8.5-9.5$. 
\cite{Huang2012} and \cite{Kreckel2012} find that specific star formation rates begin to decrease more substantially in 
galaxies with stellar masses greater than $M_*/M_{\sun}\sim10^{9.5}$, 
but this does not explain the presence of the dip in the ALFALFA optical LF.  

To estimate the ALFALFA void and wall galaxy LFs, 
we separately fit exponential cut-offs to the bright ends ($M_r\le-18$) 
and power law slopes to the faint ends ($M_r>-18$). 
The best fit characteristic magnitude in voids is $M^*= -19.95\pm 0.07$, and 
$M^*= -20.49\pm 0.04$ for walls. 
We find a shift towards fainter magnitudes in voids of 
$\Delta M\sim0.5$. This magnitude shift is much smaller than the full magnitude shift found in 
either SDSS sample, because \hi\ surveys detect bright galaxies in denser regions far less frequently 
than optical surveys.  
The separately fit faint-end slopes of the ALFALFA galaxy LFs are: 
$\alpha= -1.49\pm 0.03$ for voids and $\alpha= -1.52\pm 0.05$ for walls. 
Unlike the LFs of \sdssnearby\ galaxies over the same volume, 
we find no evidence for an environmental dependence on the faint-end slope parameter. 
This is likely because the \hi\ selected sample primarily detects late-type galaxies, whose LFs were shown 
by \cite{Tempel2011} to be independent of large-scale environment.  

We also see a much steeper faint-end slope than for the \sdssnearby\ sample. 
The effect of LSB galaxies present in the ALFALFA sample steepens the faint end of the optical LF 
closer to $\alpha\sim-1.5$ as predicted in \cite{Blanton2005ELL}. 
We believe this result is evidence that the ``true'' faint-end slope of the optical LF 
is around $\alpha\sim-1.5$. 
Thus, simulations may need to scale back the effects of feedback and photoionization 
to account for the presence of LSB galaxies.
We also note that the presence of LSB galaxies in the 
ALFALFA sample is not the cause of the intriguing dip feature in the ALFALFA optical LF.

\acknowledgments 
The authors would like to acknowledge the work of the entire
ALFALFA collaboration team in observing, flagging, and extracting 
the catalog of galaxies used in this work. 
This study was supported by NSF grant AST-1410525. 
The ALFALFA team at Cornell is supported by NSF grants AST-0607007 
and AST-1107390 to RG and MPH and by grants from the Brinson Foundation.

Funding for the creation and distribution of the SDSS Archive has been
provided by the Alfred P. Sloan Foundation, the Participating
Institutions, the National Aeronautics and Space Administration, the
National Science Foundation, the U.S. Department of Energy, the
Japanese Monbukagakusho, and the Max Planck Society. The SDSS Web site
is http://www.sdss.org/.

The SDSS is managed by the Astrophysical Research Consortium (ARC) for
the Participating Institutions. The Participating Institutions are The
University of Chicago, Fermilab, the Institute for Advanced Study, the
Japan Participation Group, The Johns Hopkins University, the Korean
Scientist Group, Los Alamos National Laboratory, the
Max-Planck-Institute for Astronomy (MPIA), the Max-Planck-Institute
for Astrophysics (MPA), New Mexico State University, University of
Pittsburgh, Princeton University, the United States Naval Observatory,
and the University of Washington.

\bibliographystyle{./apj}
\bibliography{./bibliography2}

\end{document}